\documentclass[fleqn,usenatbib]{mnras}

\usepackage{newtxtext,newtxmath}
\usepackage[T1]{fontenc}

\DeclareRobustCommand{\VAN}[3]{#2}
\let\VANthebibliography\thebibliography
\def\thebibliography{\DeclareRobustCommand{\VAN}[3]{##3}\VANthebibliography}

\usepackage{graphicx}	% Including figure files
\usepackage{amsmath}	% Advanced maths commands
\usepackage{amstext,amscd,bm}
\usepackage{array}
\newcommand\an{Astron. Nachr}

\newcolumntype{R}{>{$}r<{$}}
\newcolumntype{L}{>{$}l<{$}}
\newcolumntype{A}{R@{${}\pm{}$}L}
\newcolumntype{E}{R@{${}-{}$}L}
\newcolumntype{B}{R@{${}\,/\,{}$}L}
\newcommand{\mcl}[1]{\multicolumn{2}{c}{#1}}
\newcommand{\mcll}[1]{\multicolumn{2}{c|}{#1}}
\newcommand{\mcc}[1]{\multicolumn{1}{c}{#1}}
\newcommand{\mcu}[1]{\multicolumn{3}{c}{#1}}
\title[Photometry and dynamics of the NGC 1513]{CCD $UBV(RI)_{KC}$ photometry and dynamics of the open cluster NGC 1513}

\author[{\.I}nci Akkaya Oralhan et al.]{
{\.I}nci Akkaya Oralhan$^1$,
Hikmet {\c C}akmak$^2$,
Y\"uksel Karata\c{s}$^3\thanks{E-mail: karatas@istanbul.edu.tr}$,
Ra\'ul Michel$^4$~,
and Charles Bonatto$^5$
\\
$^{1}$Department of Astronomy and Space Sciences, Faculty of Arts and Sciences, Erciyes University, Talas Yolu, 38039, Kayseri, T\"urkiye\\
$^{2}$Department of Computer Sciences, Science Faculty, {\.I}stanbul University, 34134, Vezneciler-Istanbul, T\"urkiye\\
$^{3}$Department of Astronomy and Space Sciences, Science Faculty, {\.I}stanbul University, 34116, \"Universite-Istanbul, T\"urkiye\\
$^4$Observatorio Astron\'omico Nacional, Universidad Nacional Aut\'onoma de M\'exico, Apartado Postal 877, C.P. 22800, Ensenada, B.C., M\'exico\\
$^5$Universidade Federal do Rio Grande do Su, Departamento de Astronomia, CP\,15051, RS, Porto Alegre 91501-970, Brazil\\
}

\date{Accepted XXX. Received YYY; in original form ZZZ}

\pubyear{2024}

\begin{document}
\label{firstpage}
\pagerange{\pageref{firstpage}--\pageref{lastpage}}
\maketitle

\begin{abstract}	
We derive astrophysical parameters of the open cluster NGC~1513 by means of colour indices built with new  CCD~$UBV(RI)_{KC}$ photometry. Based on early-type members, the mean foreground reddening and total to selective extinction ratio are E(B-V)=0.79$\pm$0.09 mag and R$_{V}$=2.85$\pm$0.05. Through the differential grid method, we derive the metal abundance [Fe/H]$=$-0.06 dex (Z$=$+0.013), which is consistent with the value [Fe/H]$=$-0.088 of the bright giant member-$LAMOST-695710060$. The Z$=$+0.013 isochrone fit to the V x (B-V) colour-magnitude diagram leads to a turn-off age of 224$\pm$27~Myr (thus an intermediate-age cluster), and a distance modulus of $(V_{0}$--$M_{\rm V}$)=10.90$\pm$0.15~mag, thus implying a distance from the Sun d$=$1514$\pm$105~pc. Within the uncertainties, our photometric distance is consistent with the value d$=$1435$\pm$14~pc from the Gaia DR3 parallax. We find signs of small mass segregation through a minimum spanning tree analysis for the 190 most massive stars, together with the rather steep mass function ($\chi=$ +2.39) slope. The high core to half-light radius ratio $R_{core}/R_{h}= 0.82$, together with the compact half-light to tidal radius ratio $R_{h}/R_{t}=0.22$, suggest that it is probably related to cluster-formation effects, due to little dynamical evolution, instead of driving its dynamical evolution by internal relaxation. Indeed, NGC 1513 is located in the second quadrant ($\ell=152^{\circ}.59$ and Galactocentric distance $R_{GC}=9.57$ kpc), which tends to minimize tidal effects by external processes and tidal disruption. Therefore, internal mass segregation effects in NGC~1513 seem to be more efficient than cluster evaporation processes. We find that NGC 1513 migrated about 0.50 kpc from its birth place.
\end{abstract}


%%%%%%%%%%%%%%%%%%%%%%%%%%%%%%%%%%%%%%%%%%%%%%%%%%
%%%%%%%%%%%%%%%%% BODY OF PAPER %%%%%%%%%%%%%%%%%%

\section{Introduction}
$U$ magnitudes are of importance for deriving the reddenings, photometric metal abundances, and also ages of the open clusters (OCs). The errors of $U$-magnitudes increase towards fainter magnitudes. The errors of $\sigma_{U-B}$ are always larger, due to the smaller sensitivity of the CCD in the ultraviolet. Most CCDs which have low UV response often produce an effective wavelength too far to the red when the same photoelectric filters were used \citep{sung2000}. The so-called red leak in the used $U$ filter also worsens the situation. As emphasized by \cite{bes2005}, owing to large differences in $U$ filter between CCDs,  achiving of a good match is difficult. The $U$ and $B$ filters with Landolt system do not provide good match to the Johnson system as those used by Cousins, thus cause some systematic differences in some colour indices between the Landolt and Cousins $UBVRI$ systems \citep{lan73,lan83}. These potential effects cause a scattered distribution in colour-colour diagrams (CC), and thus  make reddening/metallicity detects difficult. In same cases such obstacles can be overcome by limiting the errors of the $(U-B)$. Due to these well-known reasons, using $U$-filter observation is discouraging for most researchers, and they prefer to use $BVRI$ filters.  In this sense NGC~1513 is a poorly studied OC in the literature with $U$-filter observation. The only study with CCD $U$-filter in the literature is from \cite{sag2022} (hereafter S22).

For deriving the reddening of the OCs via the CC diagrams, $Q$ method as the reddening-independent quantity is utilised for young  OCs with early type stars (especially B-type stars), whereas for the old-aged OCs with AFGK-type stars, a de-reddened ZAMS (Zero Age Main Sequence) fit is applied to the observations. 

In this paper we contribute to the literature for the mentioned deficiency above with the recent CCD $UBV(RI)_{KC}$ photometry of the open cluster, NGC~1513 which resides in Perseus spiral arm (Fig.~1). As stated by \cite{suc07,tap10, akk15}, since the OCs within the Sierra San Pedro M\'artir National Astronomical Observatory (SPMO, hereafter) Open Cluster Survey have been selected to be small or comparable to the size of the field of view of the CCD, the central part of each cluster can be isolated to increase the contrast of the cluster with respect to field stars in the various CC and colour-magnitude diagrams (CMD). In this respect, the SPMO survey has the advantage for identifying various groups of stars in CC and CMD plots, such as red giant/red clump (RG/RC) stars and possible Blue Stragglers. Note that RG/RC stars are also quite important for determining the ages and especially the distances of the clusters.

On the side of dynamical evolution, the members of the OCs undergo the internal and external perturbations such as stellar evolution, two-body relaxation, mass segregation, tidal interactions with the Galactic disc and bulge, spiral arm shocks, Galactic tidal field, and collisions with GMCs  \citep{Lamers2006,Gieles2007}. As stated by \cite{Tarr2022} (hereafter T22), old OCs are more mass segregated than young OCs. In the sense low-mass stars should be moved to the outer parts of the OCs,  whereas massive stars sink into their central parts. This is known as mass segregation. Internal dynamics can shape clusters cores after 100 Myr, while younger OCs  inherit their shape from the initial conditions of the cluster formation. The collisions with massive GMCs, tidal effects from spiral arms, disc and Bulge crossings are known as the external perturbations, whereas internal processes are two body relaxation and mass segregation. 

The constructed relations among the dynamical indicators  $c$, $R_{core}$, $R_{RDP}$, $R_{h}/R_{t}$, $R_{core}/R_{h}$,  $R_{h}/R_{t}$ and $R_{h}/R_{J}$, $\tau$,  $t_{rlx}$ in terms of age and  $R_{GC}$ (kpc) make it allow to interprete internal/external dynamical evolution of the OCs.  For this, some valuable studies can be given as the following, \cite{schil06}, \cite{pis07}, \cite{Camargo2009}, \cite{Bon2010}, \cite{Gunes2017} (hereafter G17), \citet{pia17a,pia17b,pia09}, \cite{Angelo2020,Angelo2021, Angelo2023} (hereafter A20, A21, A23), and T22.
See Sect.10 for the meanings of these parameters.
\begin{figure}\label{fig-1-galpos}
	\centering{\includegraphics[width=0.80\columnwidth]{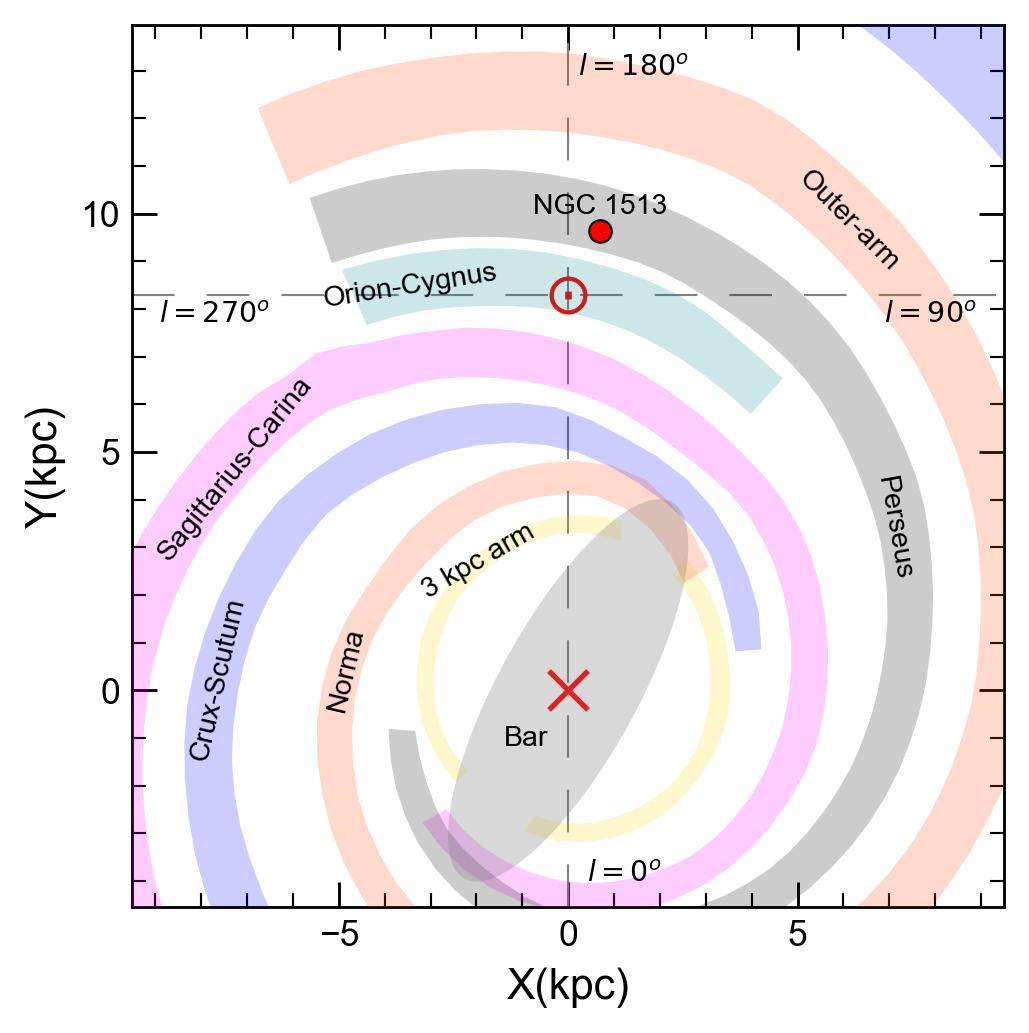} \vspace{-3mm}
	\caption{Spatial distribution of NGC~1513 (filled red  circle). 
		The schematic projection of the Galaxy with its spiral arms is seen from the North pole. $(X,~Y)$~kpc shows the galactosentric cartesian coordinates. The sun is at 8.2 kpc.}
	}
\end{figure}

In light of the information summarized above,  we obtain the colour excess, distance, and age of NGC~1513 from both CCD $UBV(RI)_{KC}$ and Gaia DR3 photometric data. The metal abundance of our sample OC for the isochrone selection is determined from the differential grid technique (hereafter DG) \citep{net13,net15,net16}. Gaia DR3 astrometric/photometric data have also been utilised for the separation of the cluster members, the obtaining of the structural/dynamical parameters, and kinematics.
By combining the findings of the astrophysical/structural/dynamical parameters, the dynamical evolution of NGC~1513 is interpreted. Radial migration is also studied from its birth radius.

This paper is organized as follows. CCD $UBV(RI)_{KC}$ photometric analysis of NGC~1513 is given in Section~2. The cluster membership technique is presented in Section~3. The derivation of reddening, distance modulus/distance, ages from both CCD $UBV(RI)_{KC}$ and Gaia DR3 photometries are presented in Sections~4-6. Structute parameters, kinematics, orbital parameters , mass/mass function slope, dynamical parameters, and radial migration are mentioned in Sections~7-10. A Discussion/Conclusion is presented in Section~11 together with a comparison with the literature.
\begin{figure}\label{fig-2-chart}
	\centering{\includegraphics[width=0.80\columnwidth]{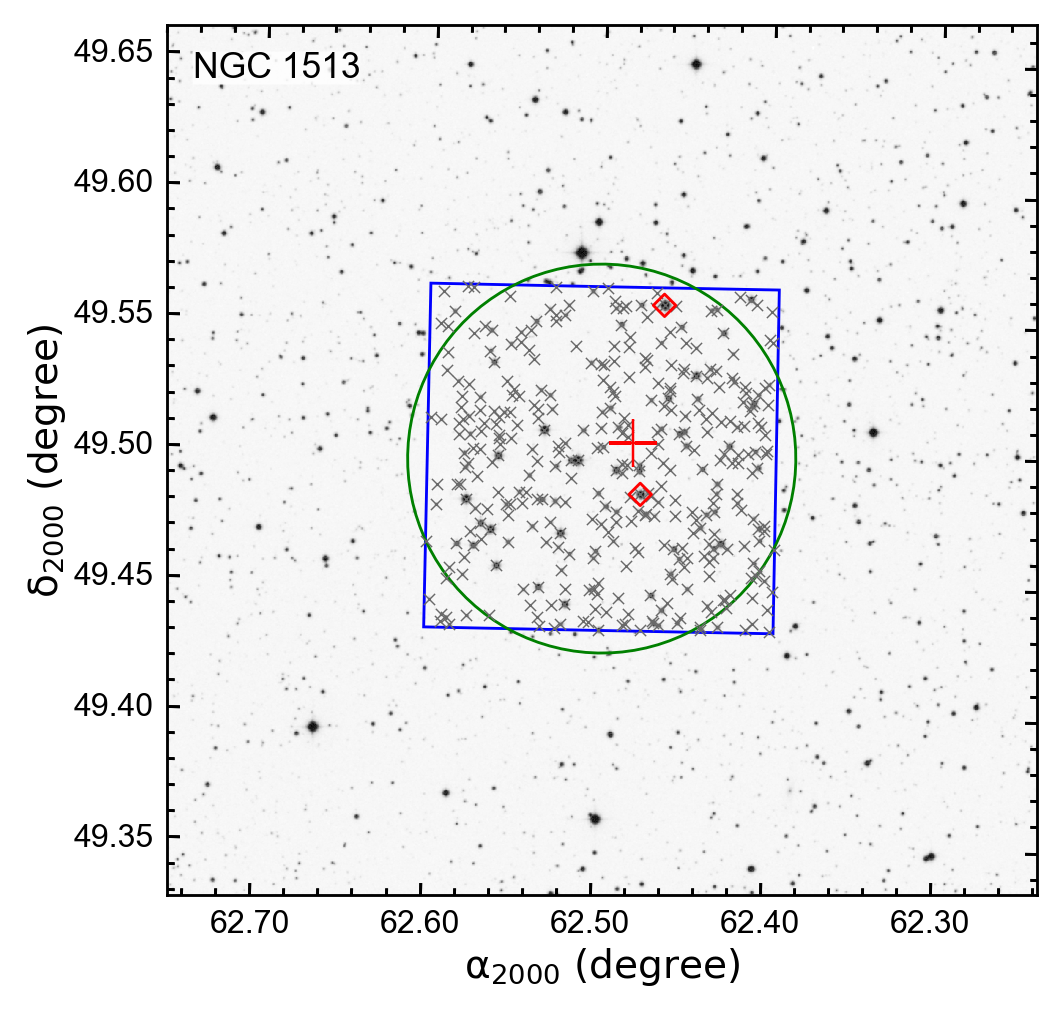} \vspace{-3mm}
		\caption{The star chart of NGC 1513 from the webpage of The STScI Digitized Sky Survey. Blue square shows the field of view of the SPM detector, $7.90^{\prime} (N-S)\times7.90^{\prime}(E-W)$ arcmin$^2$ which corresponds to  $12.350^{\prime} (N-S)\times7.887^{\prime}$ arcmin$^2$ as the lengths of the field in arcminutes. The cross, big plus and the green circle symbols represent  the cluster stars, the central equatorial coordinates (Table 1) and  core  radius (Col.~5 of Table 7), respectively. Two bright giants are shown by red diamonds (See Sect. 5).}}
\end{figure}

\section{Observation and Data Reduction}
Observation of NGC~1513 cluster and some Landolt standard fields were carried out at the SPMO during the very photometric nights of December 3-5, 2023 UT using the 0.84-m (f/15) Ritchey-Chretien telescope equipped with the Mexman filter wheel and a detector (a 2048x2048 13.5-$\mu m$ square pixels E2V CCD42-40 with a gain of 1.39 e$^-$/ADU and a readout noise 3.49 e$^-$). Binning 2$\times$2 was employed and the combination of telescope and detector ensured an unvignetted field of view of 7.90$\times$7.90 arcmin$^2$. NGC~1513 was observed through the Johnson's $UBV$ and the Kron-Cousins' $RI$ filters with short and long exposure times  in order to properly record both bright and faint stars in the region. Its star chart is displayed in Fig.~2. Standard star fields \citep{lan09} were also  observed at zenith $\approx$ 60 degrees and at meridian to properly determine the atmospheric extinction coefficients. The log of the observations in Table~1 contains the central equatorial/Galactic coordinates, air mass range during the observation, and exposure times in each filter. The flat fields were taken at the begining and end of each night, and bias images were recorded between cluster observations. Data reduction was carried out by Raul Michel with the IRAF/DAOPHOT\footnote {IRAF is distributed by the National Optical Observatories, operated by the Association of Universities for Research in Astronomy, Inc., under cooperative agreement with the National Science Foundation.} package \citep{stet87}. Standard magnitude for a given filter $\lambda$ is obtained using the following relation.

\begin{equation}
	M_{\lambda} = m_{\lambda} - [k_{1\lambda} -k_{2\lambda}C)] X + \eta_{\lambda} C + \zeta_{\lambda}
\end{equation} 

where $m_{\lambda}$, $k_{1\lambda}$, $k_{2\lambda}$, $C$, and $X$  are the observed instrumental magnitude, primary/secondary extinction coefficients, colour index and air mass, respectively. $M_{\lambda}$, $\eta_{\lambda}$, $\zeta_{\lambda}$ are standard magnitude, transformation coefficient and photometric zero point, respectively.  More details on data reduction together the extinction coefficients and zero points for $UBVRI$ filters can be found in the papers of \cite{akk10}, \cite{akk15} and \cite{akk19}. The photometric errors in $V$ and colours  $(R-I)$, $(V-I)$, $(B-V)$, $(U-B)$ against  $V$ for NGC~1513 are presented in Fig.3. The mean errors for $V$-mag intervals are listed in Table 2. Stars towards fainter magnitudes have errors larger than $\sim0.10$ mag in both magnitude and the colours. Larger errors of $\sigma_{U-B}$ as expected are seen at V $>$15 .
\renewcommand{\tabcolsep}{12mm}
\renewcommand{\arraystretch}{1.1}
\begin{table}\label{tbl-1-summary}
	\footnotesize
	\begin{center}
		\caption{Equatorial/Galactic coordinates and observation summary of NGC 1513.}
		%\vspace{-5ex}
		\begin{tabular}{lr}
			\hline
			$\alpha(2000)$\,(h\,m\,s)                              & 04 10 00.0   \\
			$\delta(2000)\,(^{\circ}\,^{\prime}\,^{\prime\prime})$ & +49 29 27.6  \\
			$\ell\,(^{\circ})$                                     & 152.59       \\
			$b\,(^{\circ})$                                        & -01.59       \\
			Airmass                                                & 1.054 – 1.060\\
			U (Exp.Time (s))                                       & 10, 100, 1000\\
			B (Exp.Time (s))                                       & 6, 60, 600   \\
			V (Exp.Time (s))                                       & 4, 40, 400   \\
			R (Exp.Time (s))                                       & 2, 20, 200   \\
			I (Exp.Time (s))                                       & 2, 20, 200   \\
			\hline
		\end{tabular}
	\end{center}
\end{table}

The photometry for NGC~1513 is compared to the $CCD~UBV$ photometry of S22. The cross-identification of the photometric catalogues of our and S22 are correctly matched by taking into consideration the difference with  $<1^{\prime\prime}$ in $\Delta \alpha$ and $\Delta\delta$. The differences of $\Delta V$ and $\Delta (B-V)$ against $(B-V)$ for 64 common stars, and the difference of $\Delta (U-B)$ against $(U-B)$ for 40 common stars are displayed in Fig.~4. The mean differences together dispersions of $V$ and $(B-V)$ are $\Delta V=+0.023\pm0.060$~mag and $\Delta (B-V)=+0.021\pm0.031$~mag, respectively. For the interval of  $0.6 < (B-V) < 1.4$, the majority of the differences of $V$ magnitudes are less than $\sim$0.05~mag, except for a few scattered stars.  $(B-V)$ values of this paper are in good consistent with S22 at level of $\Delta (B-V)=+0.021$~mag (panel~b). In panel (c) there appears a discrepancy between -0.10~mag to -0.50~mag in $(U-B)$ between our and S22 observations.  The difference $\Delta (U-B)=-0.16\pm0.17$~mag indicates that the $(U-B)$ values of SPMO are bluer than S22. For bright giant (red diamond), the discrepancy is quite small about 0.04~mag in $\Delta V$ and -0.01~mag in $\Delta (B-V)$.  $U$-magnitudes of SPMO appear to be brighter than S22. The potential reasons for large colour difference between $(U-B)$ of present photometry and S22 may be due to the used number of the standard fields, the used different effective wavelengths of $U$-filters, the different reduction techniques, and the so-called red leak in the used $U$ filter. Note that this paper and S22 use $U$ filters of Johnson's and Bessell's.

\cite{mon2023} obtained Gaia Synthetic Photometric (GSP) data-base for various photometric filters by utilising the externally calibrated fluxes of the Gaia DR3 $BP$ and $RP$. Their GSP data-base allow exquisite photometric calibration of ground-based data-sets. To see the performance of the GSP synthetic $UBVRI$ photometry for Johnson-Kron-Cousin system, we plot the magnitude differences of $UBVRI$ for the cluster members of NGC~1513 in common with the GSPC (Fig.~5). The differences between $U$-magnitudes of SPM (filled red dots) and GSP have been plotted against SPMO-$V$ mag.  $\Delta U$ difference in general fluctuate around the value $-$0.09 for $11 < V < 14$, increasing somewhat fainter magnitudes. As for the discrepancies with S22, there appears to be good concordance up to $V=14.50$ for only three stars but the differences increase towards the fainter magnitudes. For $V > 15$, SPMO-$B$ magnitudes are systematically bluer at a level of $\sim$0.03~mag relative to the GSP. The discrepancies for $V$, $R$ $I$-magnitudes are satisfactorily small with small wiggles for entire $V$-mag range. As is evident from their fig.5 of \cite{mon2023}, the sensitivity of $BP$ spectrophotometer for $\lambda < 400$ nm is low and multi-structured with two steep branches at $\sim$330 nm and $\sim$390 nm. In the sense, the $U$ magnitudes from both ground-based and space-based fluxes in ultraviolet ($UV$) region are always problematic, and thus $BP$ performance of Gaia DR3 in the $UV$ filter is not satisfactory good. Whereas $RP$ appears to give more consistence synthetic magnitudes with the observations.  
\renewcommand{\tabcolsep}{3mm}
\renewcommand{\arraystretch}{1.0}
\begin{table}
	\centering
	\caption{The mean photometric errors of $V$, $(R-I)$, $(V-I)$, $(B-V)$ and $(U-B)$ for NGC~1513 in terms of $V$ mag.}
	\begin{tabular}{cccccc}
		\hline
		\hline
		&  & &    NGC~1513  & &  \\
		\hline
		V &$\sigma_{V}$&$\sigma_{R-I}$&$\sigma_{V-I}$ &$\sigma_{B-V}$ &$\sigma_{U-B}$ \\
		\hline
		11 - 12 & 0.042 & 0.010 & 0.044 & 0.063 & 0.056 \\
		12 - 13 & 0.034 & 0.014 & 0.038 & 0.053 & 0.050 \\
		13 - 14 & 0.023 & 0.010 & 0.025 & 0.035 & 0.059 \\
		14 - 15 & 0.018 & 0.009 & 0.020 & 0.028 & 0.093 \\
		15 - 16 & 0.013 & 0.009 & 0.014 & 0.020 & 0.190 \\
		16 - 17 & 0.018 & 0.010 & 0.019 & 0.022 & 0.303 \\
		17 - 18 & 0.032 & 0.014 & 0.034 & 0.040 & 0.377 \\
		18 - 19 & 0.062 & 0.019 & 0.063 & 0.082 & 0.463 \\
		19 - 20 & 0.131 & 0.033 & 0.133 & 0.182 &  --   \\
		20 - 21 & 0.232 & 0.052 & 0.234 & 0.351 &  --   \\
		\hline
	\end{tabular}%
	\label{tbl-2-photerr}%
\end{table}%
\begin{figure}
	\centering{\includegraphics[width=0.70\linewidth]{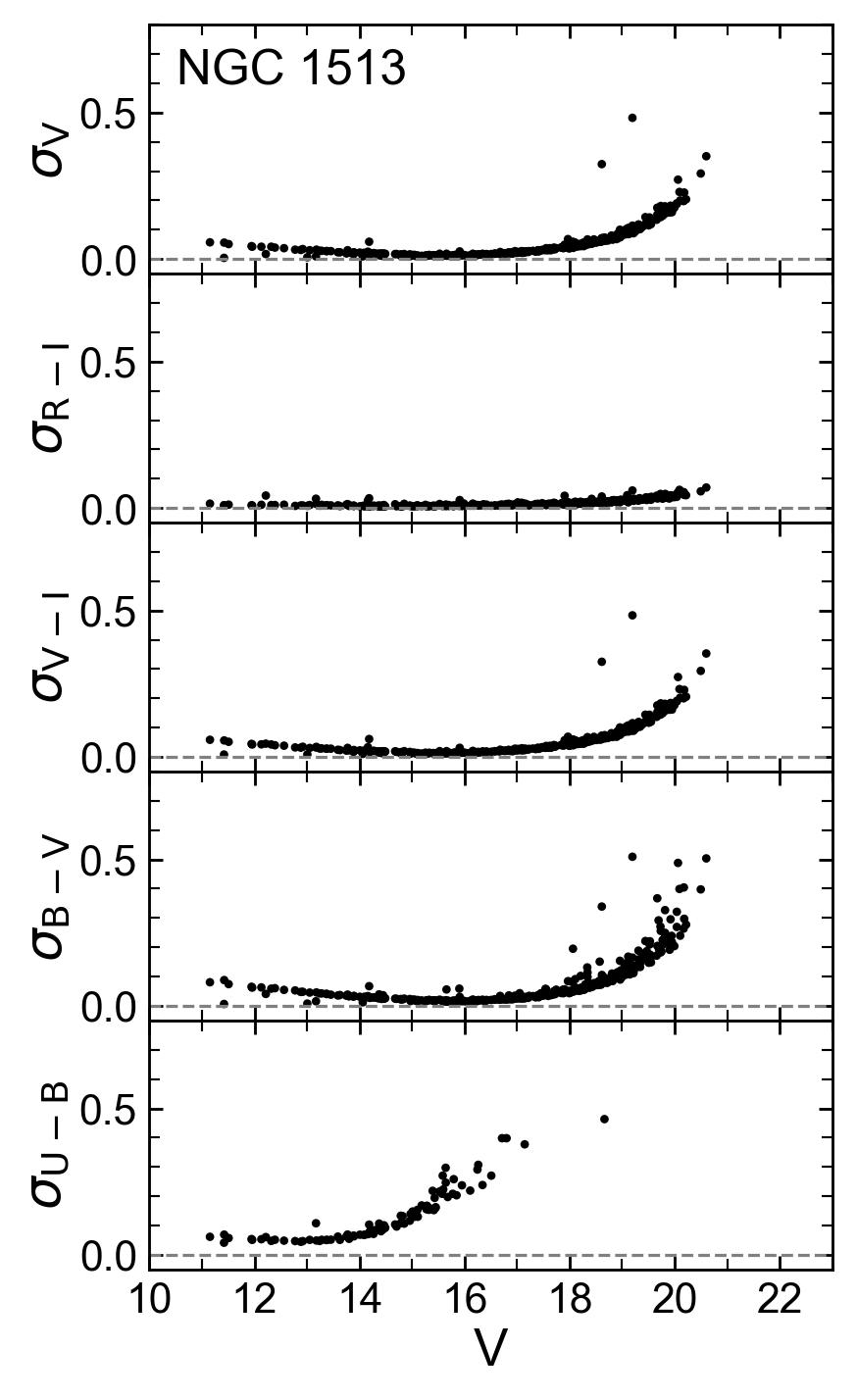} \vspace{-4mm}
		\caption{The distribution of the photometric errors of $V$, $(R-I)$, $(V-I)$, $(B-V)$ and $(U-B)$ against $V$ mag.}
	}\label{fig-3-app_pherr}
\end{figure}

\begin{figure}\label{fig-5-photocomp}
	\centering{\includegraphics[width=0.8\linewidth]{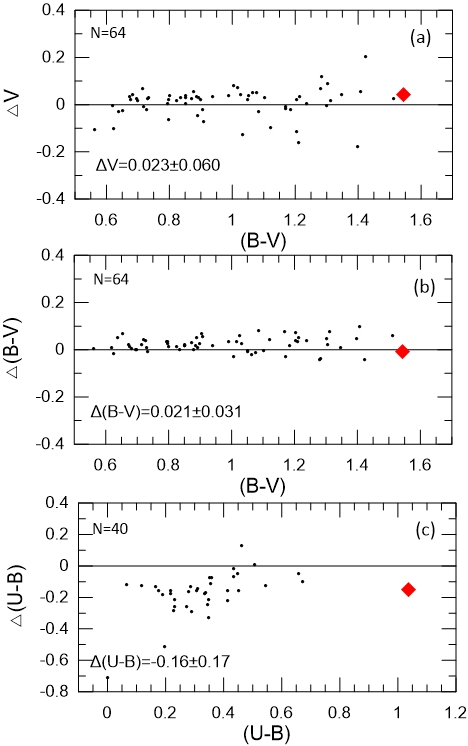}}
	\caption {The differences of $V$-mag, $(B-V)$ and $(U-B)$ as a function of $(B-V)$ and $(U-B)$. $\Delta$ means this paper -- S22. Red diamond denotes the bright giant ($LAMOST-695710060$) for just presentation (for the symbol, see also Figs.~6 and 8).}
\end{figure}

\begin{figure}\label{fig-6-diffgaia}
	\centering{\includegraphics[width=0.7\columnwidth]{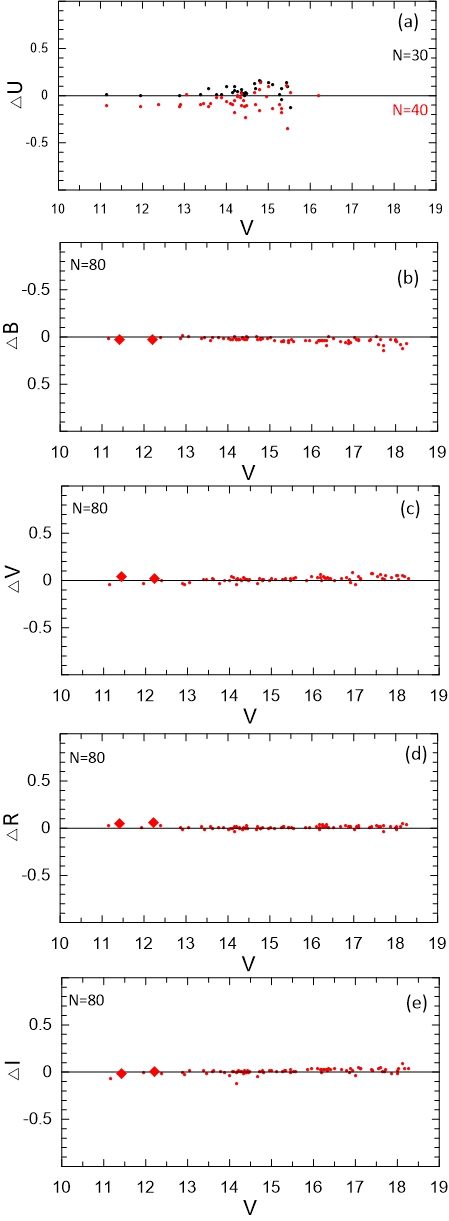}}
	\caption {The differences of the magnitudes of $UBVRI$ against SPM-V-mag. $\Delta$ means between SPMO and GSP. In panel (a), comparison with S22 is also shown with the filled black dots. Red diamonds denote two bright giants.}
\end{figure}

\section{Membership selection}
In order to separate the cluster members of NGC~1513,  its Gaia DR3 astrometric/photometric data \citep{val2023} for a large-area (15 arcmin) from VizieR\footnote{http://vizier.u-strasbg.fr/viz-bin/VizieR?-source=II/246.} are utilised. 
On the ($\mu_{\alpha}$,~$\mu_{\delta}$), the potential cluster members present a more concentrated structure, whereas field stars have a more scattered distribution.  We applied  the Gaussian Mixture Model (GMM) and SCIKIT-learn package \citep{ped11} to the cluster stars, and thus their membership probabilities P($\%$) have been determined.  The GMM\footnote{$P$ is defined $\Phi_c$ /$\Phi$.  Here $\Phi = \Phi_c + \Phi_f$ is the total probability distribution. \textit{c} and \textit{f} subscripts for cluster and field parameters, respectively. The used parameters for estimation of $\Phi_c$ and $\Phi_f$ are $\mu_{\alpha}$, $\mu_{\delta}$, $\varpi$, $\sigma_{\mu\alpha}$, $\sigma_{\mu\delta}$, $\sigma_\varpi$.} model considers that the distribution of proper motions of the stars in a cluster's region can be represented by two elliptical bivariate Gaussians. The expressions used can be found in the papers of \cite{bal98}, \cite{wu02}, \cite{sar12}, \cite{dia18},  \cite{Cakmak2021}, and \cite{Cakmak2022}. The first significant rise ($>50\%$) in the  membership probability histogram is taken into account as the membership percentage. The estimated median equatorial coordinates from the likely members are listed in Table 1. We find its median Gaia-DR3 parallax $\varpi=0.679\pm0.048$~mas. This parallax value is corrected for the Gaia DR3 parallax zero-point, $-$0.021 mas offset given by \cite{gro2021}. The distance of NGC~1513 is obtained as 1435$\pm$14 pc from the posterior probability density functions (PDFs)\footnote{https://www2.mpia-hd.mpg.de/~calj/gedr3-distances/main.html} given by \cite{Bailer2018}. The inverse parallax relation gives a distance, 1520$\pm$111 pc. However the discrepancy of both distances is quite small, 85 pc. The median proper motion components, the median parallax/bayesian distance, and the number of the members of NGC~1513 are listed in first row of Table 3. Within the errors its median values are in compatible with the one of \cite{cantat2020} (bottom panel).  

Once we assigned the cluster membership probabilites,  we cross-matched the coordinates ($\alpha$, $\delta$) of cluster stars in our ground-base photometric data with those in Gaia DR3. We set 1 arcsec as the maximum difference in ($\alpha$, $\delta$) for merging both data-sets.
\renewcommand{\tabcolsep}{1.6mm}
\renewcommand{\arraystretch}{1.3}
\begin{table}\label{tbl-3-proper}
	\centering
	\caption{The median proper motion components and  parallax/distance, and the number of  
		the probable members derived by us (top row) and Cantat-Gaudin et al.(2020) (bottom row).}
	\begin{tabular}{lccccc}
		\hline
		Cluster & $\mu_{\alpha}$ & $\mu_{\delta}$ & $\varpi$ & $d$ & N\\[-1ex]  
		&   (mas/yr)     &   (mas/yr)     &   (mas)  &  (pc)  & \\ 
		\hline
		NGC\,1513  & 1.31$\pm$0.11  & -3.70$\pm$0.10 & 0.679$\pm$0.040 & 1435$\pm$14 &482 \\
		& 1.32$\pm$0.18  & -3.68$\pm$0.16 & 0.650$\pm$0.060 & 1540$\pm$310 & 332 \\
		\hline	
	\end{tabular}
\end{table}

\section{Derivation of Reddening}
The reddening of NGC~1513 is determined from 35 early type members (relatively large red dots) with $V < 15$~mag and $\sigma_{U-B} \leq 0.10$\footnote{$\sigma_{U-B}$ and $\sigma_{B-V}$ errors of the cluster stars are generally selected to be less than $ 0\fm10$, frequently $\lesssim 0\fm05$, on the $(U-B),(B-V)$ diagram (CC) to allow us reasonable $E(B-V)$ and photometric $[M/H]$ detects.  The errors of $\sigma_{U-B}$ are also always larger than the ones of $\sigma_{B-V}$, due to the smaller sensitivity of the CCD in the ultraviolet.} on the CC (Fig.~6). $E(B-V)$ values of the members have been estimated by using the intrinsic colour relation of early type stars in tables 2-3 of \cite{sung13} (hereafter S13). The following equations of $(B-V)_{o}=0.328Q-0.019$, $Q=(U-B)-0.72(B-V)-0.025E(B-V)^2$, and $E(U-B)=0.72E(B-V)+0.025E(B-V)^2$ of S13 have been applied to the members. The median reddening is obtained as $E(B-V)= 0.79\pm0.09$~mag. Thus the reddened colour sequence of S13 (red curve) is fitted to the members (Fig.~6). The selected 35 members almost present a tight distribution around the reddened S13 main-sequence.

To test the reddening law of NGC~1513, we estimated the colour excess ratios E(V-$\lambda$)/E(B-V) of the members from the intrinsic colour relation of S13. Thus, the total-to-selective extinction ratio $R_{V}$ is obtained from  the relation between $R_{V}$ and colour excess ratios $E(V-\lambda)/E(B-V)$ of optical-near infrared colours, given by \cite{gue89}. The obtained colour excess ratios are listed in Table 4. The weighted average  $R_{V}$ is estimated as 2.85$\pm$0.05 (Table 4). This value implies that the reddening law toward NGC~1513 $(\ell=152^{o}.59,~b=-1^{o}.59)$ is slightly low as compared to 
standard $R_{V}=3.1$ and also the mean value $R_{V}=3.18$, given by  \citep{bus1978} for spectral type, $B3V-B8V$.  
However, the interstellar reddening law $R_{V}$ is known to be different for different sightlines \citep{fiz2009}.  As discussed by S13, many young OCs are known to show an abnormal reddening law. \cite{sung14} also confirmed the variation of $R_{V}$ with Galactic longitude from optical and 2MASS data of 162 young OCs (see their fig.~2). They report that the young OCs in the Perseus arm have relatively small $R_{V}$ values ($\sim2.9$). Note that NGC~1513 resides in Perseus spiral arm (see Fig.~1). 

To de-redden the two late-type bright giants, their intrinsic $(B-V_{o})$ colours are obtained by utilising the spectral type-$T_{eff}$ relation in table 5 of of S13. Their estimated reddening values are $E(B-V)=0.66\pm0.05$~mag and $E(B-V)=0.70\pm0.05$~mag, respectively (row~7 of Table~5). Thus, the giant intrinsic sequence of S13 for the two giants is reddened by amount of $E(B-V)=0.68$~mag (Fig.~6). We also determined their reddenings from table II of \cite{fer1963} as $E(B-V)=0.69\pm0.09$~mag and $E(B-V)=0.70\pm0.05$~mag. Both the two detects are in pretty compatible. 

\begin{figure}
	\begin{center}
		\includegraphics[width=0.85\linewidth] {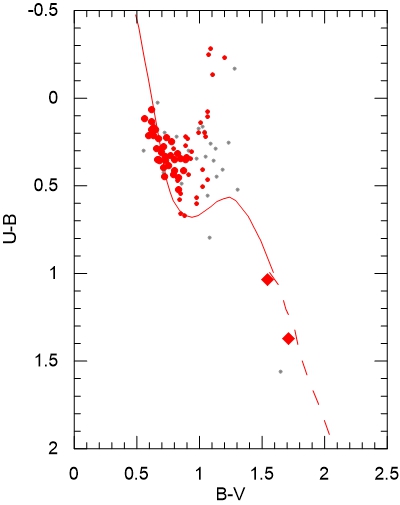} \vspace{-2mm}
		\caption{The CC diagram for 35 early -type members (large red dots) with $\sigma_{U-B} \leq 0.10$. The relatively small red and gray dots show 16 members with $\sigma_{U-B}>0.10$ and 95 member/non-members, respectively. Red solid/dashed curves, and red diamonds represent the reddened S13 dwarf/giant sequences, and bright evolved giants, respectively.}
		\label{fig-4-CCVUB}
	\end{center}
\end{figure}

\renewcommand{\tabcolsep}{6mm}
\renewcommand{\arraystretch}{1.3}
\begin{table}\label{tbl-4-color}
	\begin{center} 
		\caption {E(V-$\lambda$)/E(B-V) ratios (Col.~2) in terms of four colour indices (Col.~1). R$_{V}$ is the weighted average of four colours. Here $\lambda$ is I, J, H and K$_{s}$. N (last column): early type member numbers.}
		\begin{tabular}{lll}
			\hline
			Colour & E(V-$\lambda$)/E(B-V)  & N \\
			\hline
			\
			V-I      &1.198$\pm$0.038&35 \\
			V-J      &2.076$\pm$0.090&35  \\
			V-H      &2.404$\pm$0.10&35 \\
			V-K$_{s}$&2.574$\pm$0.15&35 \\
			\hline
			&   R$_{V}$=2.848$\pm$0.054  \\
			\hline
		\end{tabular} 
	\end{center} 
\end{table} 

\begin{figure}\label{fig-10-redmap}
	\centering{\includegraphics[width=0.90\linewidth]{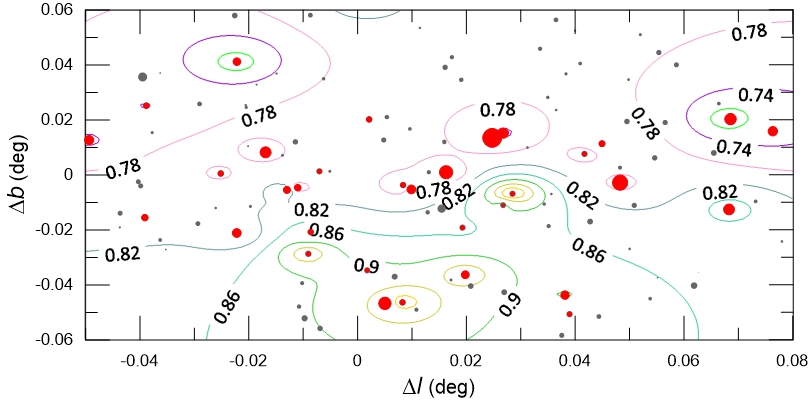}}
	\caption {Reddening map of NGC~1513. Red and gray dots shows the member and non-member cluster stars, respectively. The size of the dots is scaled to the $V$-mag of the star. The coloured line types as iso-reddening contours represent different amounts of reddening $E(B-V)$.}
\end{figure}

The spatial variation of reddening of NGC~1513 which is derived from its member stars is presented in Fig.~7. The coloured contours as iso-reddening represent different amounts of reddening $E(B-V)$. 
Judging by the map in Fig.~7 and the mean $E(B-V)= 0.79$~mag, there may be variations of $\sim$0.10~mag across the field of NGC~1513.  Although non-negligible this variation does not affect significantly the result.
Most young OCs with ages younger than 10 Myr show a differential reddening across the field of view. But for relatively old OCs, it is not well known whether there is a differential reddening across the cluster or not. That is the reason to check the differential reddening across the cluster.
\renewcommand{\tabcolsep}{1mm}
\renewcommand{\arraystretch}{1.1}
\begin{table}\label{tbl-5-gaints}
	\caption{Information of the two bright giants in NGC~1513. The photometry of Frolov et al (2002) and Sagar et al (2022) is also presented for the comparison.}
	\begin{tabular}{llll}
		\hline
		&\tiny LAMOST 695710060	&\tiny 2MASS J04095395+4929043& \\
		&NGC-1513 RH88	&NGC-1513 RH 62&Reference\\
		\hline
		$V$               & 11.420 & 12.216& This study \\ 
		& 11.400 & 12.230& F02 \\ 
		& --     & 12.173& S22 \\ 
		$(B-V)$           &1.709   &1.545& This study \\
		&1.640   &1.430& F02   \\
		& --      &1.552& S22\\ 
		$(U-B)$           & 1.372  & 1.035& This study \\ 
		& --      & 1.188& S22 \\
		\hline 
		\
		$\alpha(2000)$\,(h\,m\,s)      & 04 09 51.041   & 04 09 53.970 &\\ 
$\delta(2000)\,(^{\circ}\,^{\prime}\,^{\prime\prime})$& + 49 33 25.270 & +49 29 04.45 &\\ 
		Membership   &0.99   &0.99&\\ 
		$E(B-V)$          & 0.70   &0.66&    \\ 
		$m~(M_{\odot})$   & 3.66   & 3.45&  \\ 
		$M_{V}$           & -1.45  & -0.66& \\
		$\varpi$ (mas)         & 0.702$\pm$0.015  &0.796$\pm$0.021&\\
		$(\mu_{\alpha}, \mu_{\delta})$ $(mas/yr)$ &(+1.201, -3.736)&(+1.259, -3.648)\\	
		$V_{rad}$ (km/s)	      &-15.198$\pm$0.178&-15.18$\pm$0.39\\
		$[Fe/H]$	     &-0.088 &--\\
		$d_{plx}$ (pc)      & 1425$\pm$30       &1256$\pm$33&\\
		$d_{ph}$ (pc)       & 1496$\pm$105      &1581$\pm$101&\\ 
		\hline
	\end{tabular}
\end{table}

\section{Metal abundance and bright giants of NGC 1513}
The metal abundance of NGC~1513 has been obtained by using the DG  technique developed by \cite{poh10}, and improved by \cite{net13,net22}. The details of the DG method can also be found in the papers of \cite{net16,net22}.  For this, the photometric data of main-sequence stars of NGC~1513 have been transformed to luminosities and mean effective temperatures, the latter based on up to five colour indices using 2MASS \citep{skr06}, Gaia~DR3, and our photometric data. These have been compared to ZAMS normalised B12 PARSEC isochrones. The DG method derives its metal abundance as $[Fe/H]=-0.06\pm0.15$ $(Z=+0.013\pm0.004)$. 

The identification names of the two bright giants in Fig.~6 has been taken from SIMBAD database as  2MASS J04095395+4929043 (NGC~1513 RH62) and  LAMOST 695710060 (Gaia DR3 247308962849970560) (NGC~1513 RH88). The information of photometry, membership probability, reddening, mass/absolute magnitude, Gaia astrometry/Gaia radial velocities, metal abundance, Gaia/photometric distances are listed in Table 5. 
LAMOST database give the astrophysical parameters of the bright giant LAMOST 695710060 as $T_{eff} = 4493$ K, $\log g=2.2$, and $[Fe/H]= -0.088$. This value is in good agreement with $-$0.06 dex (the DG method). Its $T_{eff}$ and $[Fe/H]$ are also in compatible with the $T_{eff} = 4786$ K and $\log g=2.07$ which the MIST/MESA isochrone gives. 

The two giant members with high membership probabilities (row 6 of Table 5) remain within SPMO FOV (see Fig.~2). \cite{fro2002} (hereafter F02) identify these two bright giants in late-evolution stages in their CMD (see their fig.5), whereas S22 identify only one bright giant-2MASS J04095395+4929043 (NGC~1513-RH62) (see their fig.7). 
The positions of the bright giants in Fig.~11(a) coincide with the photometric data of three studies.

F02 give their absolute magnitudes as $M_{V}=-0.41$ and $M_{V}=-1.20$ with the aid of the 
isochrone of \cite{gir00}. These values are not too different from our detects (row 9 of Table 5).  
It appears that 2MASS J04095395+4929043 (NGC~1513 RH62) is a red giant in the helium-burning stage of
stellar evolution, and the other giant- LAMOST-695710060 (NGC~1513 RH88) may be a asymptotic giant (AGB)  with helium and hydrogen shell burning, depending on its position on the CMD. 
Their Gaia astrometric values are agreement with the median values in Table~3 within the error limits.  
\begin{figure}\label{fig-11-cmdubvri}
	\centering{
		\includegraphics[width=0.445\linewidth]{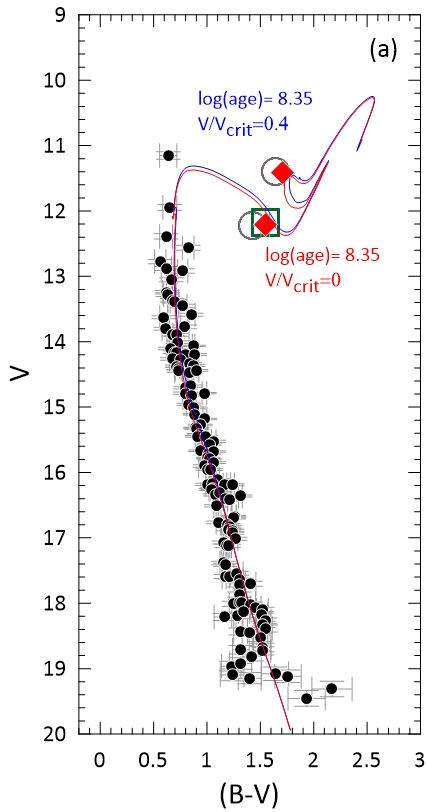}\vspace{1mm}
		\includegraphics[width=0.445\linewidth]{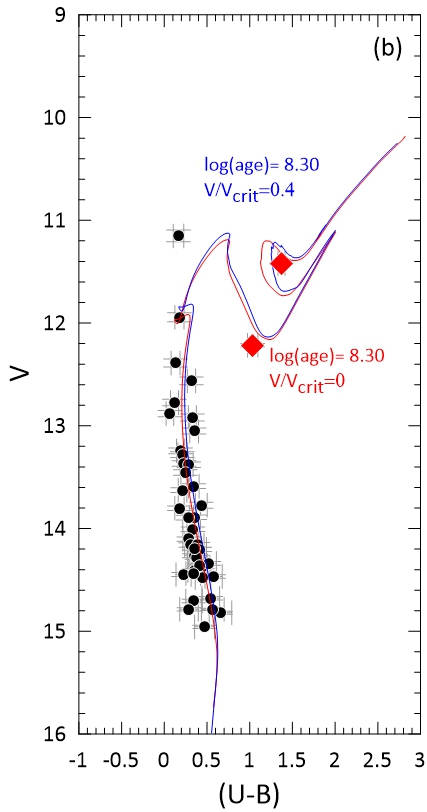}\\[0.1ex]
		\includegraphics[width=0.445\linewidth]{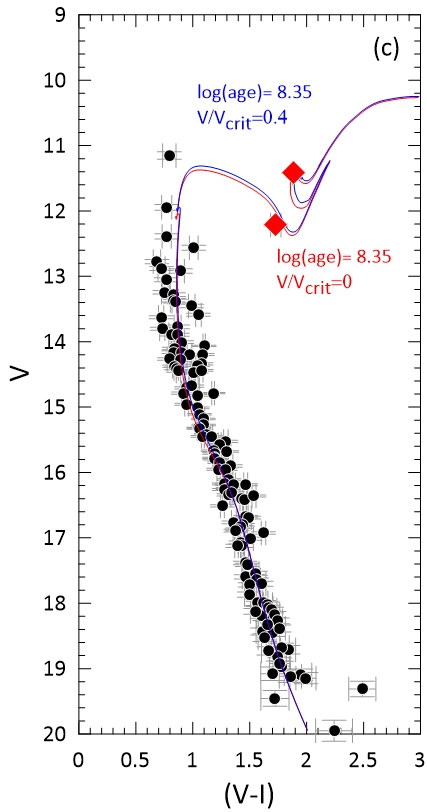}\vspace{1mm}
		\includegraphics[width=0.445\linewidth]{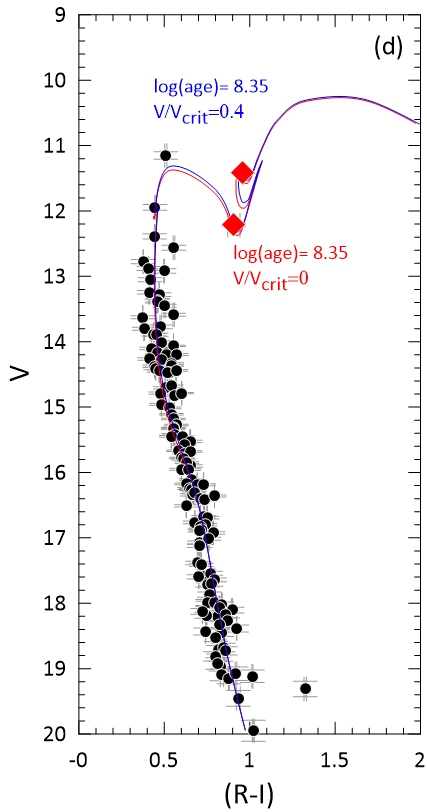}\vspace{1mm}
		\includegraphics[width=0.445\linewidth]{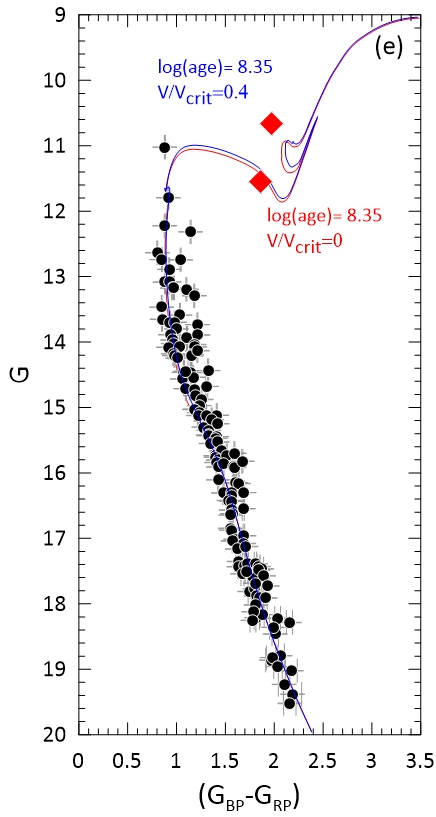}}\vspace{-2ex}
	\caption{The V-(B-V) (127 members), V-(U-B) (43 members), V-(V-I) (127 members) and V-(R-I) (127 members) and G- $(G, G_{BP}-G_{RP})$ (147 members) CMDs of NGC~1513. Red diamonds show the bright giants of this study. Large circles and a large square in upper left panel represent the ones of F02 and S22 (See Table 5), respectively.}
\end{figure}

\section{Age and distance of NGC~1513}
We used MIST isochrones (MESA Isochrones and Stellar Tracks) to derive distance modulus and age of NGC~1513. These isochones are based on the Modules for Experiments in Stellar Astrophysics (MESA) code (version v7503) \footnote[1]{https://waps.cfa.harvard.edu/MIST/interp-isos.html} \citep{cho2016,dot2016}  and \citep{pax2011, pax2013, pax2015,pax2018}. The metal abundance $([Fe/H],~Z) = (-0.06, +0.013)$ of the DG method is considered for selecting the isochrones.
Due to the diversity of the rotational velocity even in a cluster \citep{lim19}, the main-sequence turn-off broadens. The broadening in the main sequence of the CMD is also responsible for binarity, peculiarity, rotation in the stars. 
By taking into consideration this fact, the MIST/MESA isochrones with $Z = +0.013$ for both  $V/Vcrit =0.4$ (rotation) and $V/Vcrit =0$ (no-rotation) have been fitted to the five CMDs:  $V,(B-V)$, $V,(U-B)$, $V,(R-I)$, $V,(V-I)$, and $G,(G_{BP}-G_{RP})$ (Fig.~8a-e). The reddenings $E(R-I)$, $E(V-I)$, and $E(G_{BP}-G_{RP})$ have been obtained from the corresponding $E(B-V)$ value via the following relations of $E(V-I)=1.198E(B-V)$ (Table 5), $E(R-I)=0.69E(B-V)$ \citep{str1995}, and $E(B-V)=0.775E(G_{BP}-G_{RP})$ \citep{bra2018}. Note that the coefficient 1.198 is quite close to 1.25 given by \cite{dean1978}. The estimated reddenings for the colours (Col.~1) are listed in Col.~2 of Table~6.  A visual extinction of $A_{V} = 2.85\times E(B-V)$ is applied to the absolute magnitudes of the isochrones. The MIST/MESA isochrones have been shifted both vertically and horizontally to the amounts of five colour-excesses on the CMDs (Fig.~8) until the best fit to the observed intermediate section of the main sequence (MS) as well as the RG/RC sequences is obtained.  This vertical shift is the true distance modulus, $DM_{o} =(V_{0}-M_{V})$.  The derived distance moduli ($V_{0}-M_{V}$) and distances d~(pc) for five colours are given in Cols.~3-4 of Table 6. For the determination of the age, the MIST/MESA isochrones have been shifted both vertically and horizontally in the CMD's with $M_{V}+2.85E(B-V)+DM$ and $C_0(\lambda_1-\lambda_2)+E[C(\lambda_1-\lambda_2)]$, respectively. The derived ages are  given in Col.~5 of Table~6. The MIST/MESA isochrones with rotation/no-rotation closely match the observational data in the CMDs (Fig.~8), and provided consistent ages and distances (Table~6).  We report an age of $224\pm27$ Myr from  $V-(B-V)$ for NGC~1513. 
\renewcommand{\tabcolsep}{1.4mm}
\renewcommand{\arraystretch}{1.2}
\begin{table}\label{tbl-6-param}
	\centering
	\caption{Derived astrophysical parameters of NGC~1513. CE means the colour excess.}
	\begin{tabular}{lAAAA}
		\hline
		NGC~1513  &    \mcl{CE}   & \mcl{$(V-M_{V})_{o}$} &  \mcl{d(pc)} & \mcl{Age (Myr)}  \\
		\hline
		(B-V)     &  0.79&0.08  &   10.90&0.15   & 1514&105  &  224&27  \\
		(U-B)     &  0.58&0.06  &   10.90&0.20  & 1514&140  &  200&82  \\
		(V-I)     &  0.96&0.11  &   10.90&0.15   & 1514&105  &  224&58   \\
		(R-I)     &  0.55&0.06  &   10.90&0.20   & 1514&140  &  224&58  \\
		$(G_{BP}-G_{RP})$ & 0.95&0.11 &10.90&0.15  & 1514&105 &224&58  \\
		\hline
	\end{tabular}
\end{table}

The main sequence strips or bands in the CMDs are affected by the contamination of binary stars, particularly the mid-points due to stellar binaries. For this reason, we have fitted MIST/MESA isochrones to the blue- and  faint-most side of the observed sequences whenever possible, assuming that it reflects the single-star distribution \citep{car2001}.

As we experienced in \cite{akk10}, the uncertainties of distance moduli and ages of our sample OC have been derived with the help of two MIST/MESA  isochrones on the CMDs. 

For the comparisons with the literature, we have considered  mostly the astrophysical  parameters given by $(B-V)$ colour index, because the astrophysical parameters of the  OCs are mostly given, and best represented, in terms of the CMD: $V-(B-V)$.  For this,  a summary of the astrophysical parameters and the comparison to the literature for our sample OC are listed in Table~10.

The distance modulus/distance of NGC~1513 are also obtained by utilising the reddening-independent quantities, $Q'$, $Q_{V\!I}$, $Q_{V\!J}$, $Q_{V\!H}$ and $Q_{V\!K_{S}}$ from 35 early type stars. These quantities are nearly free from metallicity differences. The ZAMS of S13 in the $Q_{V\lambda}$--$Q'$ plots should be fitted to the lower ridge of the main-sequence band to avoid the effects of multiplicity and evolution, as stated  by \cite{lim14}. The ZAMS line is shifted up and down in the $Q_{V\lambda}$--$Q'$ plots by 0.1 mag. The error of this method is about 0.20 mag. Once the adjustment of ZAMS above and below the distribution of the members, its distance modulus from four colour indices is obtained as $(V_{0}-M_{V})=10.90\pm0.20$~mag, equivalent to 1514$\pm$140~pc. Within the error limits this distance falls in the distance range of Gaia DR3 parallax (Table 4) and $V-(B-V)$ (Table 6).

\renewcommand{\tabcolsep}{1.7mm}
\renewcommand{\arraystretch}{1.2}
\begin{table*}	
	{\footnotesize
		\begin{center}
			\caption{Structural parameters of NGC 1513.}    
			\label{tbl-7-struct}\vspace*{-4mm}
			\resizebox{1.0\textwidth}{!}{ 
				\begin{tabular}{lcAAAA|AAAAccc}
					\hline
					\mcc{Cluster} & $(1')$   & \mcl{$\sigma_{0K}$} & \mcl{$\sigma_{bg}$} & \mcl{$R_{core}$}
					& \mcll{$R_{RDP}$} & \mcl{$\sigma_{0K}$} & \mcl{$\sigma_{bg}$} & \mcl{$R_{core}$}
					& \mcl{$R_{RDP}$} & \mcl{$R_{t}$} & CC \\
					& ($pc$) & \mcl{($*\,\prime^{-2}$)}&\mcl{($*\,\prime^{-2}$)}
					& \mcl{($\prime$)}& \mcll{($\prime$)} &\mcl{($*\,pc^{-2}$)}& \mcl{($*\,pc^{-2}$)}& \mcl{($pc$)}
					& \mcl{($pc$)} & \mcl{($pc$)} & \\
					\mcc{($1$)} & ($2$)    & \mcl{($3$)} & \mcl{($4$)} & \mcl{($5$)} & \mcll{($6$)}
					& \mcl{($7$)} & \mcl{($8$)} & \mcl{($9$)} & \mcl{($10$)}&\mcl{($11$)} &($12$) \\
					\hline
					%%          (1')    S0K err    Sbg  err    Rcor err    Rrdp err    S0K  err  Sbg  err   Rcor err     Rrdp err     CC
					NGC\,1513 & 0.439  & 7.02&2.58 &4.63&0.16 &4.46&0.93 & 11.26&0.42 &3.08&1.13 &2.02&0.07 & 1.96&0.41 & 4.94&0.19 & &10.65& 0.956\\
					\hline
					\hline 
					
					&      &   \mcl{}  &  \mcl{}  &  \mcl{}   &   \mcll{}  &  \mcl{}  &  \mcl{} & 1.41&0.74 & 3.50&1.80& & & 1 \\
					
					&      &   \mcl{}  &  \mcl{}  &  \mcl{}   &   \mcll{}  &  \mcl{}  &  \mcl{} & 1.43&0.21 & 6.68&0.67 & & & 2 \\
					&      &   \mcl{}  &  \mcl{}  &  \mcl{}   &   \mcll{}  &  \mcl{}  &  \mcl{} & 1.65&0.26 & 6.51&0.20 & & & 3 \\
					&      &   \mcl{}  &  \mcl{}  &  \mcl{}   &   \mcll{}  &  \mcl{}  &  \mcl{} & 2.00&0.37 &\mcl{} &&5.48$\pm$0.70& 4 \\
					&      &   \mcl{}  &  \mcl{}  &  \mcl{}   &   \mcll{}  &  \mcl{}  &  \mcl{} & \mcl{1.94} &\mcl{}  & &10.63 & 5 \\		
					\hline
				\end{tabular}
			}
		\end{center}
	}
	\flushleft
	Notes:  ($*\,\prime^{-2}$) and ($*\,pc^{-2}$) in Cols.~3-4 and 7-8 represent $stars~arcmin^{-2}$ and $stars~pc^{-2}$, respectively. For the radius in pc we adopt the Gaia distance (Table~6). $R_{t}$ in Col.~11 represents the tidal radius and Col.~12 means the correlation coefficient of the RDP fit. The first row lists our results, the subsequent ones literature data - References: 1: \cite{mac2007}, 2: \cite{buk2011}, 3:\cite{Gunes2017}, 4: \cite{kha13}, 5: \cite{hunt2024}. 
\end{table*}

\section{Structural parameters}
The stellar radial density profile (RDP) of NGC~1513 is constructed from the Gaia DR3 photometric/astrometric data for cluster field within $15'.0$ down to $G =19$ mag \citep{Arenou2018}. 
The $R_{core}$, $\sigma_{bg}$, and  $\sigma_0$ of NGC~1513 have been derived by fitting two-parameter model of  \cite{King1966} to the observational RDP (Fig.~9). The two-parameter function \citep{King1966} is given as below.
\begin{equation*}
\sigma(R) = \sigma_{bg} + \frac{\sigma_0}{1+(\frac{R}{R_{core}})^2}
\end{equation*}

For this equation, $\sigma_{bg}$ is the residual background density. $\sigma_0$ and $R_{core}$ are the central density of stars and the core radius, respectively. The cluster radius $(R_{RDP})$ by comparing the RDP level with background and by measuring the distance from the centre is determined \citep{Bon2007}. The $R_{RDP}$ is taken as an observational truncation radius, whose value depends both on the radial distribution of member stars and the field density.  The large uncertainty within  $R < 1'$ in the RDP of NGC 1513 is owing to its low star content in its central part.
The three-parameter model of \cite{King1962} describes well the outer parts of a cluster, while the two-parameter model of \cite{King1966}  explains the central region of the clusters, as discussed by \cite{Bon2005}. Its tidal radius $R_{t}$ has been obtained as 10.65 pc by applying the three-parameter King function to the observational RDP (Fig.~9). These structural parameters and their meanings together with the literature values are listed in rows~2-11 of Table~7.

Our $R_{t}$ is in good agreement with the value of \cite{hunt2024}, which is based on Gaia DR3 data. The value of \cite{kha13} are smaller than our value.  As emphasized by T22,  the obtaining of the tidal radii of the OCs is quite difficult, owing to both the loose nature of the OCs and the contamination by field stars.
Our ($R_{core}$,~$R_{RDP}$) (pc) falls within or are close to the values  of \cite{mac2007}, \cite{buk2011}, G17, \cite{kha13}, and \cite{hunt2024} within the uncertainties.  

The density contrast parameter ($\delta_{c}$) to estimate the compactness of the OCs is obtained from the relation,
\begin{equation*}
\delta_{c} = 1 +\frac{\sigma_0}{\sigma_{bg}}
\end{equation*}

The $\delta_{c}$ of NGC~1513 is estimated as 2.52 from the information of Cols.~3-4 of Table 7. This value lies outside the range  $7 \le \delta_{c} \le 23$ for the compact OCs  given by \cite{Bon2009}. This means that NGC~1513 ($b= -1^{\circ}.59)$ has a sparse OC. Note that different Galactic latitudes have different properties and also suffer from differences in density contrast, since low-Galactic latitude OCs tend to be projected against denser fore/backgrounds.
In this context, \cite{Bica2005} also state that  the $\delta_{c}$ values of the OCs depend on the Galactic latitude than the cluster mass. Accordingly, the populous and high Galactic latitude OCs have large density contrast parameters.  
\begin{figure}
	\centering{\includegraphics[width=0.95\linewidth]{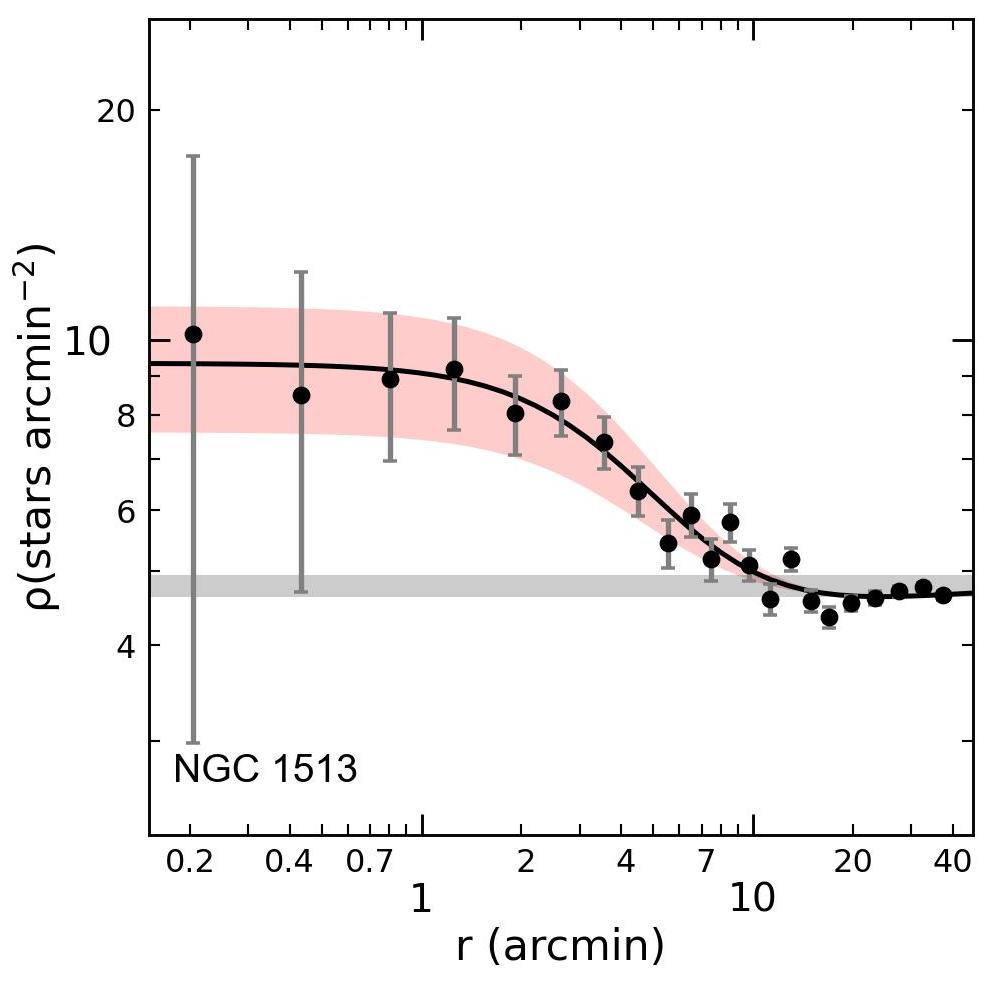} \vspace{-2mm}
		\caption{The radial density profile of NGC~1513. Solid line and horizontal gray bar represent the best fit King profile and the stellar background level measured in the comparison field, respectively. The $1\sigma$ King fit uncertainty is shown by the shaded domain.}
	}\label{fig-13-rdp}
\end{figure}

\section{Kinematics and Orbital Parameters}
The Gaia radial velocities (Table 5) of two bright giants of NGC~1513 have been taken from \cite{Soubiran2018,Soubiran2019}. The average radial velocity of the two bright giants has been assumed as NGC 1513's radial velocity.  Its heliocentric velocity ($U$, $V$, $W$) in the right-hand system has been obtained from its average radial velocity, the median proper motion components/distance of NGC~1513  by using the algorithm of \cite{joh87}. Here the photometric distance of Gaia DR3-CMD (Table~6) is adopted. Its space velocity is converted  to the components $U'$, $V'$, $W'$ by correcting for the Solar motion $(U, V, W)_{\odot} = (+11.10, +12.24, +7.25)$ km s$^{-1}$ with respect to the local standard of rest (LSR) \citep{sch10}. Here, $R_{\odot}=8.2\pm0.1$ kpc \citep{bg16} and $V_{LSR}$ = 239\,km\,s$^{-1}$ \citep{bru11} are adopted. The heliocentric cartesian distances ($x'$, $y'$, $z'$) (kpc) and LRS-velocity components ($U'$, $V'$, $W'$) km s$^{-1}$ are converted to the Galactic Rest of Frame (GSR) i.e., ($x$, $y$, $z$) (kpc) and ($V_{x}$, $V_{y}$, $V_{z}$) km s$^{-1}$. The azimuthal velocity  ($V_{\Phi}$) (km s$^{-1}$) is estimated via 
\begin{equation*} 
	V_{\Phi} =  \frac{x V_{y} - y V_{x}}{R}
\end{equation*}
Here, $V_{\Phi}<0$ means prograde.  From the "MWPotential2014" code in the galpy-code library \footnote[1]{http://github.com/jobovy/galpy} written by \cite{bov15}, peri and apo-galactic distances $(R_{min},~R_{max})$ (kpc) and the maximum height distance (z$_{max}$) (kpc) are obtained. The orbital eccentricity (ecc) is estimated via the relation
\begin{equation*} 
	ecc = \frac{R_{max}-R_{min}}{R_{max}+R_{min}}
\end{equation*}

The current mean galactosentric radius $R_{m}$ is estimated from $(R_{min}+R_{max})/2$, which is also known as the guiding or mean orbital radius. The orbits have been integrated using kinematical parameters (Table~8) for its age (Table~6) within the galactic potential. The galactic potential as the sum of the galactic components is explained by \cite{bov15}. Its orbital angular momentum component  $J_{z}$ (kpc km s$^{-1}$) is calculated from the equation  $J_{z}=x V_{y}-y V_{x}$. All these parameters are listed in Table~8.

According to rotational velocity $V_{\Phi}= 222$ km s$^{-1}$, the eccentricity, 0.06, and orbital angular momentum value (Table~8), NGC~1513 reflects the characteristics of the Galactic disk. The x-y~(kpc) plane is known as projected on to the Galactic plane, whereas z-R (kpc) is the meridional plane. On the x-y~(kpc) plane (Fig.~10), it follows a circular path around the Galactic center. According to its revolution period (T$=$250 Myr) around the Galactic center (Table~8), it has almost made one revolution around the center of the Galaxy.  The gray circles in the panels show its five future revolutions. 

The orbits in the z-R (kpc) plane show boxy-like type properties. NGC~1513  moves in the meridional planes within the confined spaces and is oscillating along the z-axis.  NGC~1513's orbit is confined in the range of $\sim 8.5 < R_{gc} \leq 9.6$ kpc. Its initial $(t = 0~Gyr)$ and present day position in our Galaxy are shown in Fig.~10 with the filled blue and red dots, respectively. These positions are calculated by using the "MWPotential2014" in the galpy-code package. For this, time is taken as zero (initial) and cluster age (present day). Its closest approaches to the Sun is at  $(d-kpc,~t-Gyr) = (0.46, 0.014)$.

\newcommand{\hpm}{$\,\pm\,$}
\renewcommand{\tabcolsep}{2.8mm}
\renewcommand{\arraystretch}{1.2}
\begin{table}\label{tbl-8-overall}
	\begin{center}
		\caption{The average radial velocity ($V_R$) km s$^{-1}$, space velocity components and rotational velocity  ($U$, $V$, $W$, $V_{\Phi}$) km s$^{-1}$, eccentricity (ecc) and peri and apogalactic distances, initial and present day distances (R$_{max}$, R$_{min}$, R$_{m}$, z$_{max}$, R$_{in}$, R$_{GC}$) (kpc). The orbital angular momentum ($J_{z}$) (kpc km s$^{-1}$). $T$ is the time of one revolution around the Galactic center as Myr. $N_{Rev}$ is the number of the revolution over the age of the cluster (last row).}
		\begin{tabular}{lcccc}
			\hline
			& NGC\,1513 & \\
			\hline
			$V_R$         & -15.38\hpm0.41 &   \\ 
			$U$           &    2.57  &   \\
			$V$           &   -29.22  &   \\ 
			$W$           &   -12.59  &   \\ 
			$V_{\Phi}$    &  -222.42  &   \\ 
			ecc           &     0.06  &   \\
			R$_{min}$     &     8.47  &    \\
			R$_{max}$     &     9.57  &    \\
			R$_m$         &     9.04  &     \\
			z$_{max}$     &     0.08  &     \\
			R$_{in}$      &     8.50  &      \\
			R$_{GC}$      &     9.57  &   \\
			$J_z$         & -2127.53 & \\
			$T$           &   250     &    \\
			$N_{Rev}$     &   1   &     \\
			\hline
		\end{tabular}
	\end{center}
\end{table}
\begin{figure}\label{fig-14-orbit}
	\centering{\includegraphics[width=0.75\linewidth]{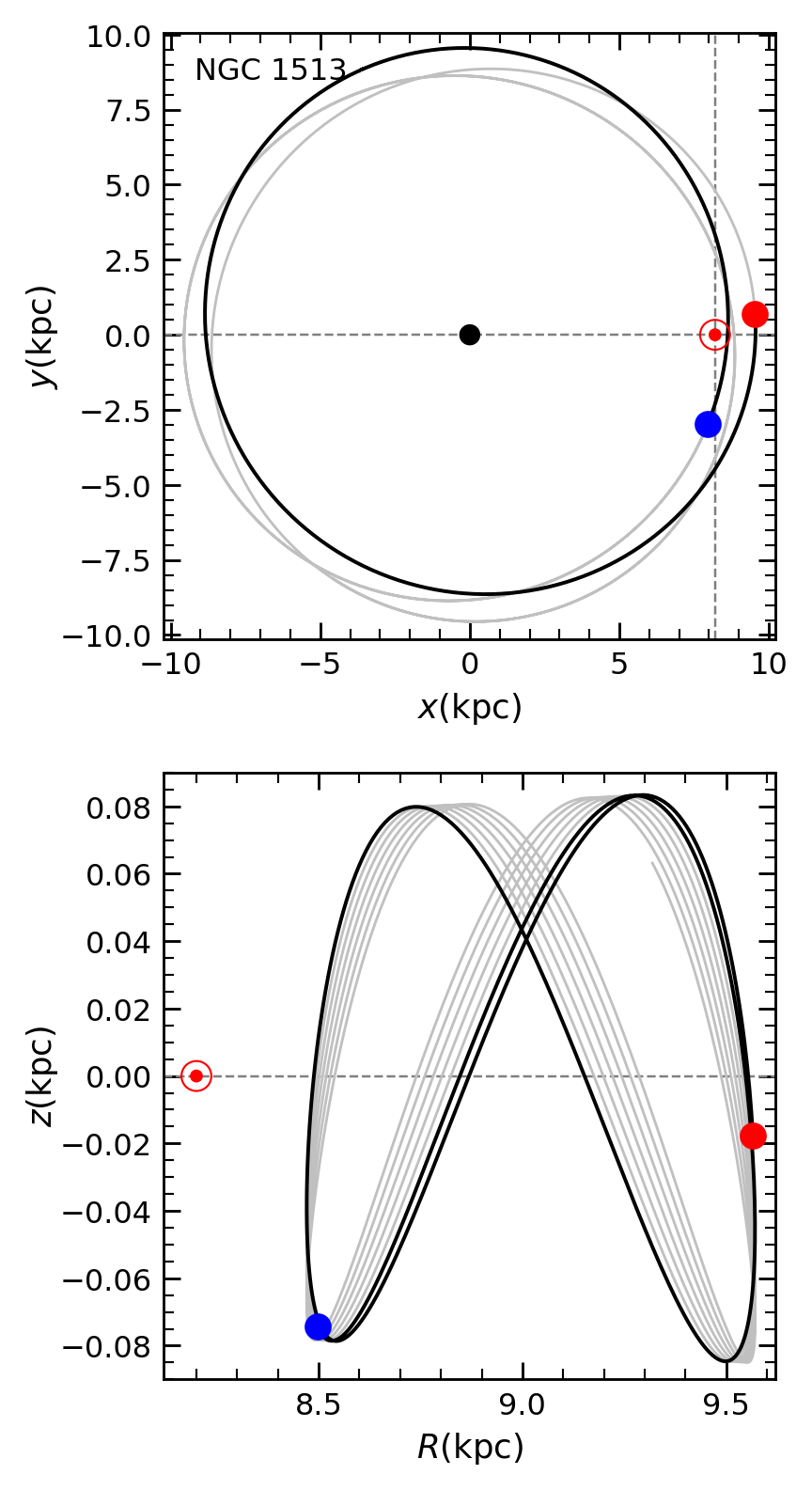}\hspace{0.5mm}}\vspace*{-3mm}
	\caption {Galactic orbit of NGC~1513. The trajectories represent the paths traveled by the OC through its age. The filled blue/red dots show its initial/present day position. The open red circle shows the Sun's position.}
\end{figure}

\section{Mass and Mass Function Slope}
The masses of the main sequence members of Gaia DR3 for NGC~1513 have been determined from the MIST/MESA isochrones. $G$ magnitudes of main sequence members have been converted into the absolute magnitudes from its  reddening and distance modulus (Cols.~2 and 3 of Table~6).  Mass function (MF) is defined as the distribution of masses of cluster stars per unit volume. The mass function in terms of 0.05 bin sizes is given in Fig.~11. The MF slope is fitted to a power-law given by,
\begin{equation}
\begin{aligned}
	\log\frac{dN}{dM}=-(1+\chi)\log(M) + \text{constant}\\[1ex]
\end{aligned}
\end{equation}
\noindent
Here $dN$ is the number of cluster members for the  mass interval $dM$ with central mass $M$, and $\chi$ is mass function slope. Since Gaia data ($G$ mag) is complete below $G=19$ mag \citep{Arenou2018},  stars brighter than this limit are taken, which correspond to stars more massive than 1 $M_{\odot}$. Applying Eq.~(2) to the observations (Fig.~11) gives NGC 1513's  MF slope as  $\chi =2.39\pm0.25$. This MF slope is very steep as compared to  $\chi=1.3\pm0.7$ of \cite{Kroupa2001} with the uncertainties. Total mass is obtained from the MF slope of NGC~1513. The mass range, MF slope, total mass, mean mass, and member number are listed in Col.~1 of Table~9.  Our MFs is steep as compared to $\chi=1.55\pm0.20$ of \cite{mac2007}, $\chi=+1.90\pm0.12$ of G17 and $\chi=1.53\pm0.37$ of S22. The relatively small mass values, $\sim625~M_{\odot}$ of this study and $713~M_{\odot}$ of G17 are almost compatible. 
\begin{figure}\label{fig-15-massfunc}
	\centering{\includegraphics[width=0.75\linewidth]{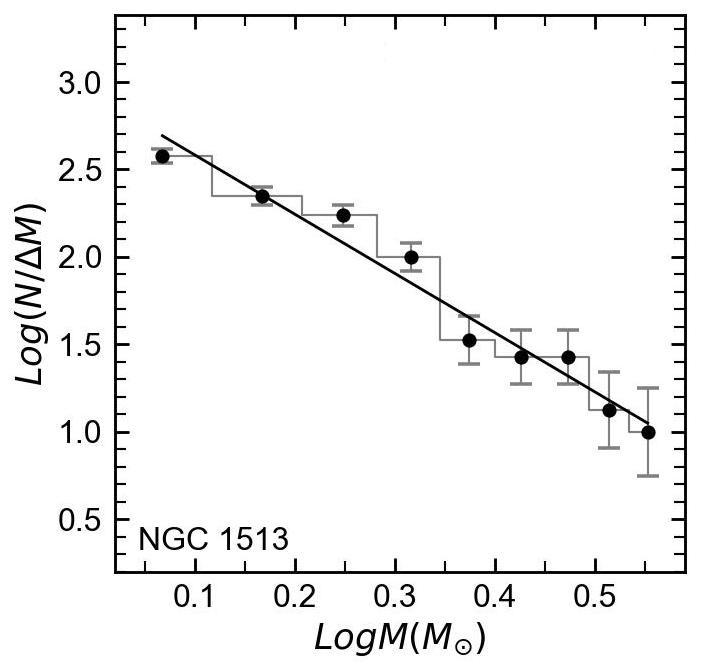}}\vspace*{-3mm}
	\caption {The mass function of NGC~1513. The vertical error bars as poisson noise are proportional to the  $\sqrt{N}$.}
\end{figure}

By following the procedure given by \cite{al2009a} and T22, we estimated the mass segregation ratio $\Lambda_{MSR}$ from Eq.(3) by using the masses of the members of Gaia DR3 of NGC~1513.  

\begin{align}
\begin{aligned}
\Lambda_{MSR} (N) = \frac{<l_{random}>}{l_{massive}} \pm \frac{\sigma_{random}}{l_{massive}},\\[1ex]
\end{aligned}\label{eq:MSR}
\end{align}

The degree of mass segregation $\Lambda_{MSR}$ is the ratio of the lengths of the
average randomly selected star (MST) to the most massive star, as explained by  \cite{al2009a}.
MST is defined as the length of the minimum spanning tree of the most massive stars of a cluster \citep{al2009a}. $<l_{random}>$ is the average length of the MST of N randomly chosen stars, and thus $<l_{massive}>$ is the length of the MST of the N most massive stars. $\sigma_{random}$ is the standard deviation of $<l_{random}>$. 
The length of the MST of the most massive stars ($l_{massive}$) is shorter than the average length of the MSTs of the random stars ($<l_{random}>$) \citep{al2009a}. In this context, the massive stars have more concentrated distribution, hence the cluster is mass-segregated. 
\cite{al2009a} state that $\Lambda_{MSR}=1$, $\Lambda_{MSR} > 1$ and  $\Lambda_{MSR}<1$ indicate no-mass segregation, mass segregation and inverse-mass segregation, respectively. 

The relation of $\Lambda_{MSR}$ versus $N_{MST}$ is displayed in Fig.~12 for NGC~1513. 
For $N_{MST} < 10$, 10 most massive stars with $\Lambda_{MSR} < 1$ show inverse mass segregation. This method places 50 most massive stars in the inner parts the cluster. Almost 190 most massive stars with $\Lambda_{MSR}>1$ are mass segregated.  This does necessarily mean that these massive stars are widely spaced than the other stars in the cluster. Almost 190 most massive stars with $\Lambda_{MSR}>1$ are mass segregated. 
First 40 most massive stars are $\Lambda_{MSR}\sim1.7$ times closer to each other in the inner radius. 
For $50 < N < 120$, $\Lambda_{MSR}$ decreases to $\sim1.35$.  $\Lambda_{MSR}$ drops to unity for around $N_{MST}=190$. Beyond this limit, no any sign of mass segregation is seen.
\begin{figure}\label{fig-16-mst}
	\centering{\includegraphics[width=0.98\linewidth]{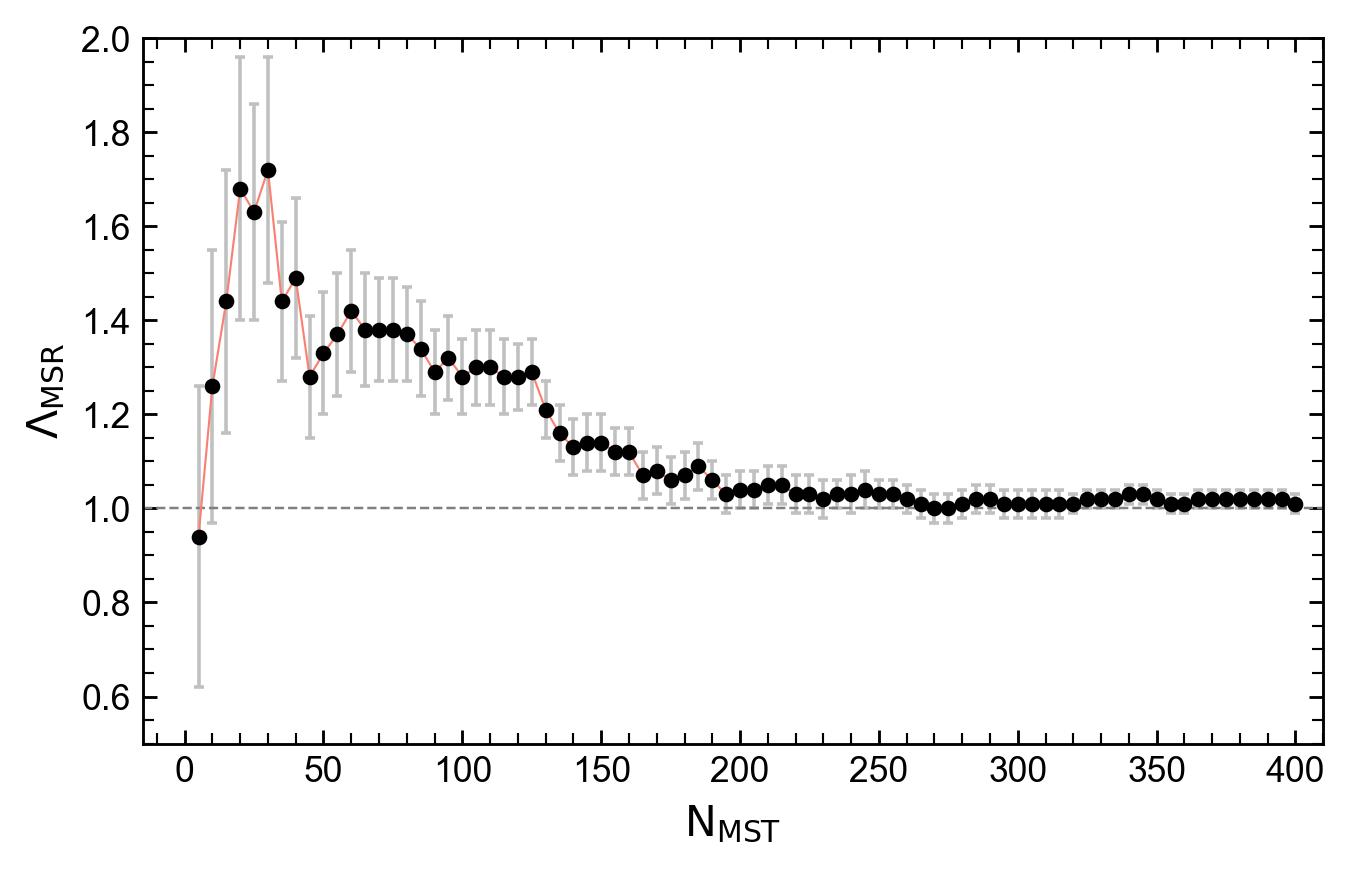}}\vspace*{-3mm}
	\caption {Mass segregation ratio $\Lambda_{MSR}(N)$ for NGC~1513 
		as a function of the N more massive stars. The horizontal red dotted line denotes the limit $\Lambda_{MSR}=1$ which means no-mass segregation.}
\end{figure}
 
\renewcommand{\tabcolsep}{3.3mm}
\renewcommand{\arraystretch}{1.2}
\begin{table*}\label{tbl-9-dynamic}
	\begin{center}
		\caption{The obtained  parameters of mass information, dynamical parameters/dynamical indicators, and migration parameters for NGC~1513.} 
		\begin{tabular}{lA @{}p{6mm}@{} | lA @{}p{6mm}@{} | lr @{}p{6mm}@{} | lr}
			\hline
			\mcu{Mass information}               & &      \mcu{Dynamical parameters}    & &   \mcl{Dynamical indicators}   & & \mcl{Migration parameters} \\
			\hline
			Mass range \,($M_{\odot}$) &   \mcl{1.12 $-$ 3.98}   & & $R_h$\,(pc)      &    2.39&0.26    & & $c$               & 0.74       & & $R_{guide}\,(kpc)$  & 9.57 \\
			MF slope \, ($\chi$)       &       2.39&0.25         & & $R_J$\,(pc)      &    \mcl{19.88}   & & $R_{core}/R_{h}$  & 0.82       & & $R_{birth}\,(kpc)$  & 9.11 \\
			Total mass \,($M_{\odot}$) &      624.7&7.4          & & $t_{rlx}$\,(Myr) &     22.3&5.9    & & $R_{h}/R_{t}$     & 0.22       & & $d_{mig}\,(kpc)$    & 0.46 \\
			Mean Mass \,($M_{\odot}$)  &       1.37&0.06         & & $t_{diss}$\,(Myr)&      421&89     & & $R_{h}/R_{J}$     & 0.12       & &                     &      \\
			Members \, (N)             &        457&21           & & $\tau$           &       11&4      & & $R_{t}/R_{J}$     & 0.54       & &                     &      \\
			\hline
		\end{tabular} 
	\end{center} 
\end{table*}

\section{Dynamical parameters}
The relaxation time $t_{rlx}$ is obtained from the relation by \cite{Spitzer1971}
\begin{equation}
\begin{aligned}
t_{rlx}=\frac{8.9\times10^5\sqrt{N}\times{R_{h}}^{3/2}}{log(0.4N)\times \sqrt{m}}\\[1ex]
\end{aligned}
\end{equation}
\noindent where $m$ and  $N$ refer to the mean mass and the number of the cluster members, respectively, taken  from Col.~1 of Table~9. $R_{h}$ is defined as the radius from the core that contains half the total mass of the cluster. This parameter has been inferred from Fig.~13, showing the cumulative mass in dependence of the radius. The half-mass radius ($R_{h}$) of NGC~1513 is 6$^{\prime}$.16.  For the conversion into pc, we considered 0.439 value (Col.~2 of Table~7). As an indicator of the dynamical evolution, the evolutionary parameter is estimated from the relation $\tau = Age/t_{rlx}$ by using its age (Gaia DR3) (Col.~5 of Table~6).  Its Jacobi tidal radius $R_{J}$ is estimated from the equation given by \cite{por2010},

\begin{equation}\label{eq_jacobi}
\begin{aligned}
R_{J}=\left(\frac{GM}{2\omega^{2}}\right)^{1/3}~~~~~~~~~
\omega =\frac{V_{c}}{R_{GC}}\\[1ex]
\end{aligned}
\end{equation}
\noindent where $\omega$, $M$, $V_{c}$, $R_{GC}$ are orbital angular frequency in the galaxy,  the observed total mass (Col.~1 of Table~9), circular velocity (row~5 of Table 8), galactosentric distance (row~12 of Table~8), respectively. The concentration parameter is estimated from the relation $c = \log (R_{t}/R_{core})$. See sect.6.1 of \cite{kar2023} for the role of dynamical indicators in understanding of the dynamical evolution of the OCs. All the obtained dynamical parameters are given in Cols.~2-3 of Table~9.

The birth radius ($R_{birth}$) of NGC~1513 is estimated  with the help of the current metallicity gradient based on young OCs, the model by \cite{min18} for the time evolution of the Galactic  Interstellar Medium (ISM) metallicity gradient, its metal abundance (-0.06 dex) and age (Gaia DR3) (Table~6). Its guiding radius and migration distance are obtained from the relations $R_{guide}=J_{z}/V_{\Phi}$ and $d_{mig} = R_{guide}-R_{birth}$, respectively. The obtained parameters are listed in the last line of Table~9.  
\begin{figure}
	\centering{\includegraphics[width=0.7\linewidth]{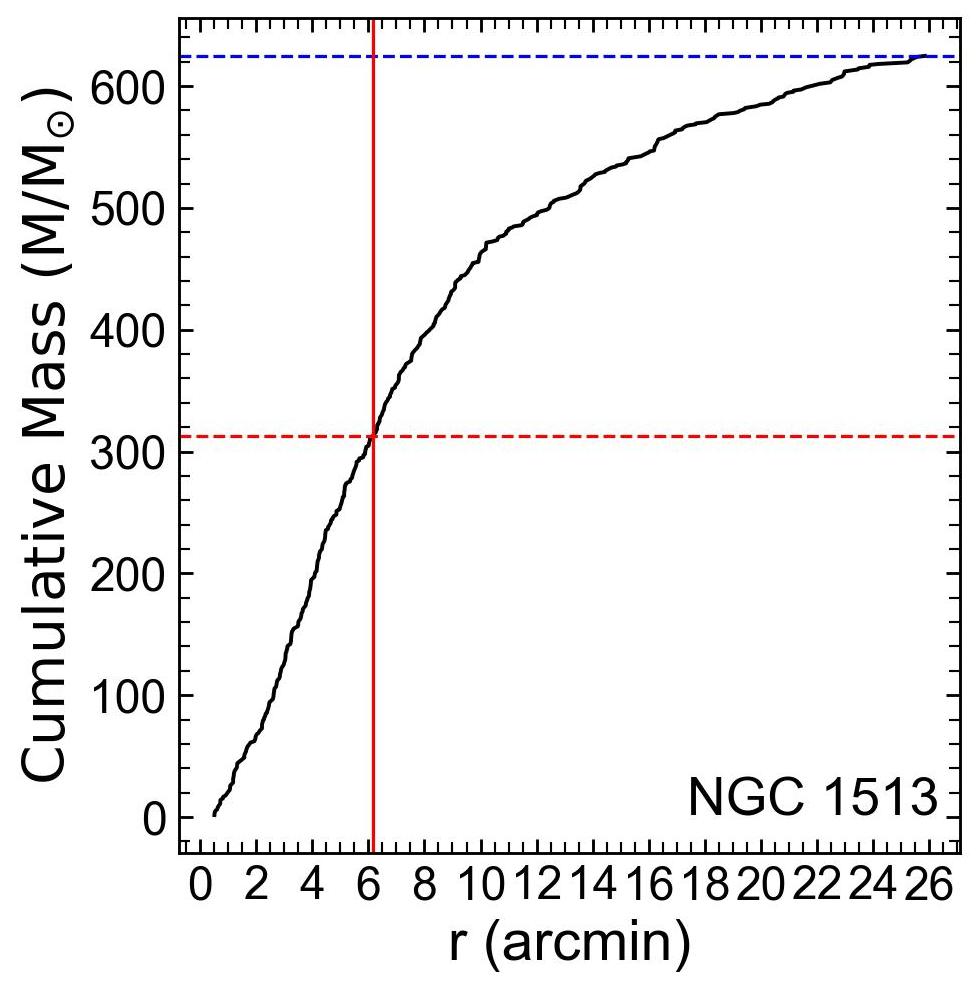}\vspace*{-3mm}
		\caption{The cumulative mass versus radius of NGC 1513. The vertical line represents the half-mass radius corresponds to the half-mass (horizontal dotted line). The upper blue dotted line denotes the total mass.}
	}
\end{figure}

\renewcommand{\tabcolsep}{0.8mm}
\renewcommand{\arraystretch}{1.2}
\begin{table*}\label{tbl-10-litcompbig}
	\caption{Literature comparison. Our results in row 1 are given for only $(B-V)$ colour index. The reddenings of different photometries are also listed in parentheses.}
	\resizebox{1.0\textwidth}{!}{\tiny%
		\begin{tabular}{llcccclll}
			\hline
			E(B-V)  & $(V-M)_{0}$& d~(pc)  & Z  & [Fe/H] & Age~(Myr)  & Isochrone  & Photometry    & Reference \\
			\hline
			NGC 1513      &   &  &   &   &    &    &        & \\
			\hline
			0.79$\pm$0.09 & 10.90$\pm$0.15 &1514$\pm$105& 0.013  & -0.06 & 224$\pm$27 & Paxton et al. (2011)  & CCD UBVRI & This paper \\
			0.65$\pm$0.03 & 10.58$\pm$0.16 &1330$\pm$100& 0.019  & --   & 225$\pm$25  & Girardi et al. (2000) & CCD UBVRI &1\\
			0.74$\pm$0.06 \tiny($E(J-H)=0.23$) &11.53$\pm$0.24 & 1180$\pm$150& 0.019  & --   & 100$\pm$20  & Marigo et al. (2008)  & JH$K_{s}$ &2 \\
			0.67 &10.95 & 1400& 0.019  & --   & 316  & Marigo et al. (2008)  & JH$K_{s}$ &3 \\
			0.76$\pm$0.16 & 10.60$\pm$1.08 & 1320$\pm$700& solar  & --    & 251   & Bertelli et al. (1994)  & CCD BV  &4 \\
			0.67 & 10.60$\pm$0.00 & 1320& 0.019  & --   & 254   & Girardi et al. (2000)  & CCD BV & 5\\
			0.70 \tiny($E(G-R)=0.93$) & 13.10 & 1323$\pm$100& solar  & --  & 150  & ZAMS- SK1965 & photographic RGU &6 \\
			0.52 \tiny($A_{v}=1.60$) & 10.95 & 1546 & solar & -- & 290 &  Bressan et al. (2012) & Gaia DR2 &7 \\
			\hline
		\end{tabular}
	}
	\\ [2ex]
	\renewcommand{\tabcolsep}{1mm}
	\renewcommand{\arraystretch}{0.8}
	\small
	\begin{tabular}{r l m{1cm} r l}
		1 &Sagar, R. et al., 2022, J. Astrophys. Astr. & & 5 &Frolov, V.N. et al., 2002,  A\&A,  396, 125     \\
		2 &Gunes, O. et al., 2017, AN, 338, 464  & & 6 &del Rio G., Huestamendia G. 1988, A\&AS, 73, 425 \\ 
		3 &Kharchenko N.V. et al., 2013, A\&A, 558, 53 & & 7 &Cantat-Gaudin, T. et al.,  2020, A\&A, 640, 1  \\
		4 &Maciejewski, G. and Niedzielski, A., 2007, A\&A,  467, 106 & &  & \\	
	\end{tabular}
\end{table*}

\section{Discussion and Conclusion}
Within the error limits our $E(B-V)=0.79\pm0.09$~mag is in concordance with the literature (Table~10). 
The discrepancies in the reddenings are quite small, $\Delta E(B-V)=$ 0.03 to +0.14, except for the value 0.52 of \cite{cantat2020}. Our distance value $(1514\pm105)$~pc (Col.~3) is good agreement with the ones of $1471\pm15$~pc (Table~3, Gaia DR3 parallax) and 1546~pc of \cite{cantat2020} (Table~10) within the uncertainties. Note that the previous literature provides close distances. The differences for distance are up to level, $\Delta d(pc)=$ 32 to 334 pc.

With the age $224\pm27$~Myr ($\log Age=8.35\pm0.05$), NGC~1513 is an intermediate age OC, according to the age interval 100$-$700 Myr ($7.0 \le \log Age < 8.85$) given by S13. \cite{del1988} and \cite{Gunes2017} give a relatively young age up to $\sim$ 70-120 Myr in terms of our age value. The age given by \cite{kha13} is somewhat older than our detect. Our age value falls in the age interval of the rest literature with the uncertainties. The discrepancies of the age are at the level of 65 to 151~Myr. 

The obtained ages from both this study (Table 10) and S22 are good agreement, although the different isochrones are used. For the isochrone selection,  this study uses $Z=+0.013$ value, whereas  S22 consider the solar abundance value, $Z=+0.019$. However, S22 give a somewhat low reddening $E(B-V)= 0.65$~mag and a close distance. This can be explained that our $(U-B)$ colours are bluer than the ones of S22 (Fig.~4c). In addition to this, as is seen from Fig.~14, the CC plot of S22 includes some early spectral-type stars as well as a few scattered FG-type stars around the bump. However, the CC plot of this study is occupied by early spectral type stars. S22 apply Schmidt-Kaler \citep{sch82} ZAMS fit to the their observations by assuming  $R_{V}=3.1$. Our $E(B-V)$ and $R_{V}$ determination is based on the $Q$ method for early type members and usage of S13 main-sequence.

\begin{figure}\label{fig-19-cccomp}
	\begin{center}
		\includegraphics[width=0.9\linewidth] {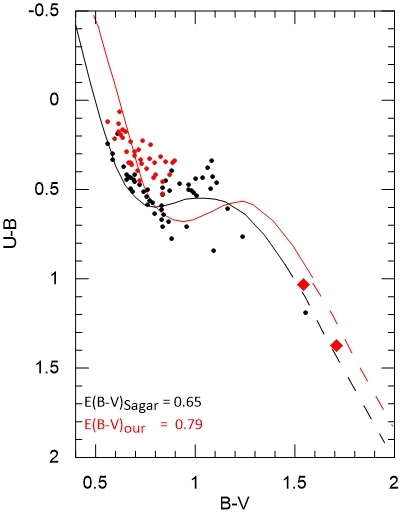}
		\caption{The comparison of the CC diagrams of NGC 1513 for our (filled red dots) and S22  (filled black dots). The curves show the reddened S13 (red) and SK82 (black)  main sequences. Red diamonds denote the bright evolved giants.}
	\end{center}
\end{figure}

The small/large discrepancies in ages and distance moduli/distances (Table~10) stem from the adopted isochrones in terms of solar abundance, the reddening values from the CMDs or CCs, as well as different reduction techniques, as emphasized by \cite{kar2023}.

The dynamic evolution of NGC~1513 has been interpreted with the help of dynamical parameters of  the literature (Figs.~15-16). The values of  $2000~M_{\odot}$ and $(R_{core},~R_{RDP})=(1.5,~7)$ pc given by G17 are considered to   classify the OCs as less massive/massive and large/small sized. 
With the $(R_{core},~R_{RDP})=(1.96,~4.94)$ pc (panel a), NGC~1513 follows the increasing trend between $R_{core}$ and $R_{RDP}$ of \cite{Camargo2009} and G17. T22 state that $R_{core}$ values of old age OCs are on average smaller than young ones. Internal dynamics can shape clusters cores after 100 Myr, while the shapes of young OCs originate from  the initial conditions of the cluster formation. Our $R_{core}=1.96$ pc value also falls in 1-2.5~pc interval in fig.7 of T22. In this context its relatively large $R_{core}$ and small $R_{RDP}$ may be due to its initial formation from molecular cloud.

Its age (224 Myr) is higher than its relaxation time ($t_{rlx}$) (Table~9). Therefore, it is dynamically relaxed.  With its overall steep MFs $(\chi=+2.39)$ and low $\tau = 10$,  NGC~1513 falls in the range of $0.50 < \chi < 2.5$ in the $\chi$ versus $\tau$,  given by  G17 (See their fig.27a). For the OCs with  $\tau>30$,  their $\chi$ values tend to be negative because of the loss of low-mass stars, whereas their MFs show a flattening for the $\tau<30$. This does necessarily mean that NGC~1513 presents a sign of small scale mass segregation with little dynamical evolution. Large scale mass segregation means that all stars in an OC have completed mass segregation, and thus the OC loses its low-mass content to the field. In the sense NGC~1513 did not lost much its low-mass stars to the field, owing to the internal/external perturbations. Its low mass stars outnumber its high mass stars, according to its MFs. Time scale parameters $(t_{rlx},~\tau)=(176~Myr,~0.57)$ of G17 also support the small mass segregation. 

NGC~1513 presents multiple levels of mass segregation (Fig.~12).  $\Lambda_{MSR}$ drops to unity for around $N_{MST}=190$. Therefore, almost 190 most massive stars with $\Lambda_{MSR}>1$ are mass segregated. Beyond this limit, no any sign of mass segregation is seen.  Accordingly, NGC~1513 ($\log Age=8.35$) falls in the interval of $\log Age$  distribution with mass-segregation of T22 (See their fig.14). 

The position of NGC~1513 in the $R1$ and $R3$ regions (Figs.~15b-c) does not indicate signs of expansion and shrinkage. The relationship between the cluster dimensions and age is linked to survival and dissociation rates of the OCs \citep{Camargo2009}. NGC~1513 resides in the second quadrant ($\ell=152^{\circ}.59$) and $R_{GC} = 9.57$ kpc (outside the solar circle) together with its relatively small mass ($\sim625~M_{\odot}$).   
\begin{figure}
	\centering{\includegraphics[width=0.93\linewidth]{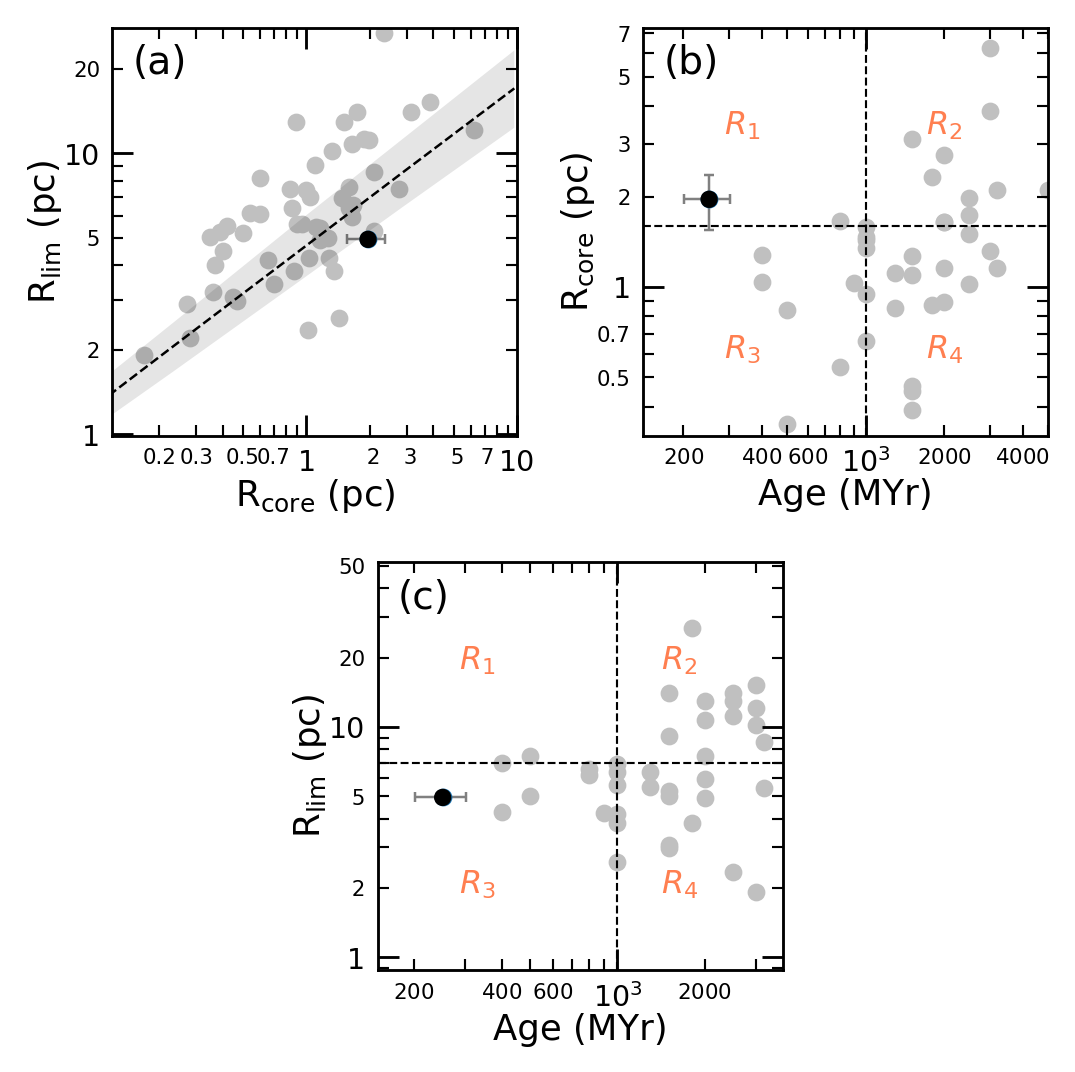}}\vspace*{-3mm}
	\caption {$R_{RDP}$ versus $R_{core}$ (panel~a),  $(R_{core},~R_{RDP})$ versus Age (panels~b-c). The relation (dotted line) and its 1$\sigma$ uncertainty as the shaded area in panel~(a), the regions $R1-R4$ in panels~(b)-(c),  the horizontal/vertical dotted lines to separate small/large sized OCs, and the comparison OCs (gray dots) are from G17.}
	\label{fig-20-keplerdyn1}	
\end{figure}
\begin{figure}
	\centering{\includegraphics[width=0.93\linewidth]{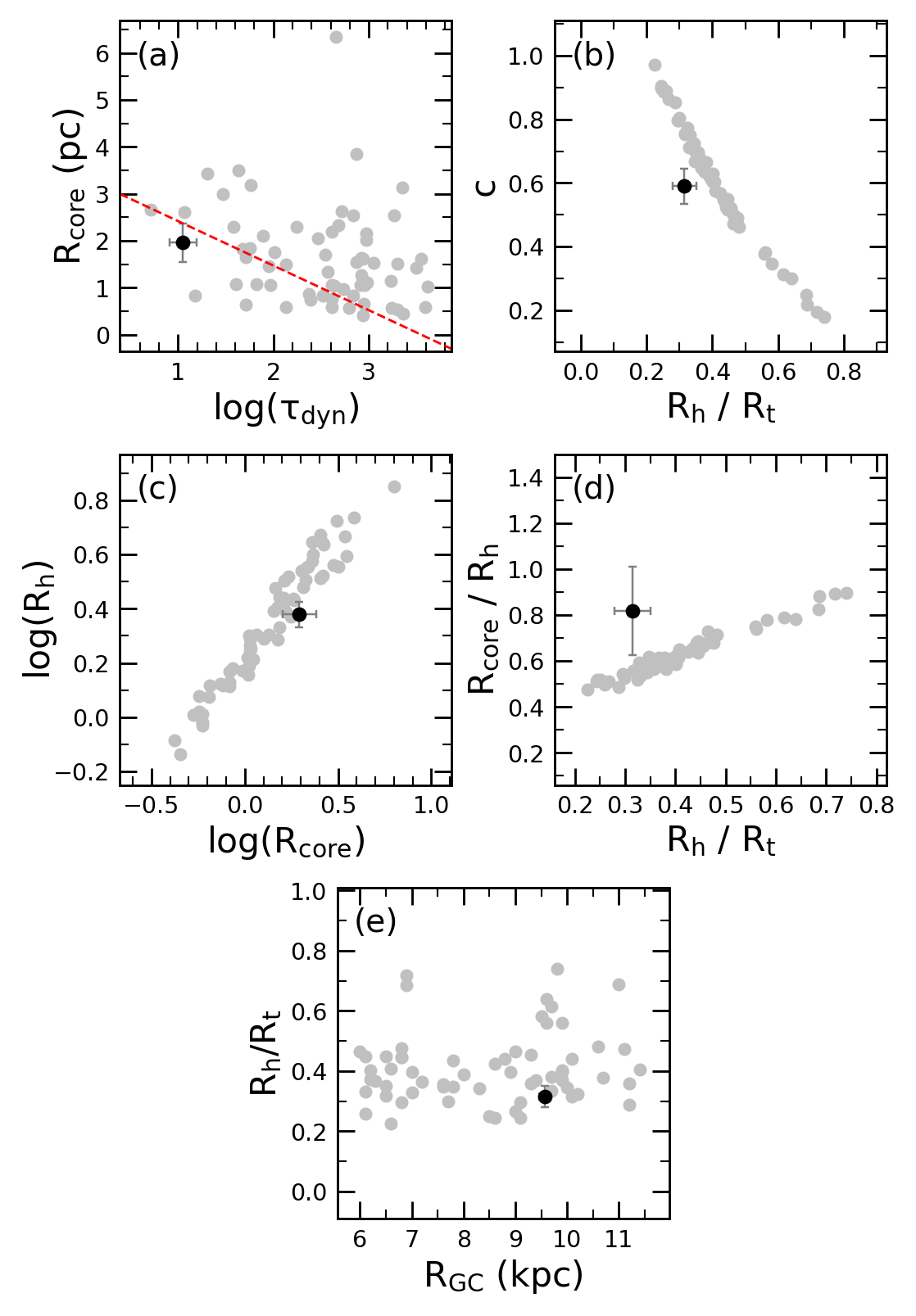}}\vspace*{-3mm}
	\caption {$R_{core}$ versus $\log(\tau)$ (panel~a), $c$ versus $R_{h}/R_{t}$ (panel~b), $\log (R_{h})$ versus $\log (R_{core})$ (panel~c), $R_{core}/R_{h}$ versus $R_{h}/R_{t}$ (panel~d), 
		$R_{h}/R_{t}$ versus $R_{GC}$ (panel~e). The gray filled dots in the panels and the dashed line (panel~a) are from A20 and A21. The position of NGC~1513 in the panels is shown with the black dot.}
	\label{fig-21-keplerdyn2}	
\end{figure}

The relations among the internal and external dynamical evolution parameters (Table~9)  
for NGC~1513 (black dot) are shown in the panels of Fig.~16 together with the data of A20-21 (gray dots).  Within the error limits NGC~1513 is almost close to the general trends in the panels. 
From panel (a), its relatively large $R_{core}$ indicate that it is dynamically evolved ($\log\tau=1.04$). Its $R_{h}=2.34$ pc is greater than its $R_{core}=1.96$~pc, and thus it meets the requirement of compliance with the general trend (panel c). 

NGC~1513 with a high $R_{core}/R_{h}=0.82$ and a compact $R_{h}/R_{t} = 0.22$ deviates from the trend in panel~(d). Note that $R_{h}$ detection between this study (Fig.~16) and A20/A21 is different, and their  $R_{h}$ is based on the relation $R_{h} = 1.3~a$ by using Plummers' $a$ parameter \citep{plummer1911} from the RDP profile fit.

Its relatively high $R_{core}/R_{h}$ ratio is probably related to cluster-formation effects due to its little dynamical evolution, instead of exposing to two-body relaxation and mass segregation in its core region.
Based on the simulations of \cite{heg2003}, the $R_{core}/R_{h}$  tends to decrease as the cluster dynamically evolves, since internal relaxation combined with the preferential evaporation of lower-mass regulated by the external potential, leads to a higher degree of central concentration. Since NGC 1513's central structure with the high $R_{core}/R_{h}$ ratio  occupies a considerably high fraction of the cluster main structure, despite the fact that it is little dynamically evolved OC ($\tau = 10$), this inflated central structure may be consequence of the formation process. Without the firm arguments based on the simulations, we avoid the possibility of rapid expulsion.

The concentration parameter ($c$) gives some evidence regarding the OC dynamical state. Its $c=0.74$ (panel b),  small $R_{h}/R_{t}=0.22$ and $R_{h}/R_{j}=0.12$ ratios imply that NGC~1513 is a compact OC. 
$R_{h}/R_{t}$ is also as an indicator of the tidal influence of the Galaxy on the dynamical evolution of the OCs. 
its $R_{h}/R_{t}$ is smaller than the tidally limit, $R_{h}/R_{t} = 0.40$ given by \cite{heg2003}. According to these compact ratios, its dynamical evolution is driven by its internal relaxation. 
Its low  $R_{h}/R_{j}$ ratio also falls in the range of 0.10-0.23 according to figure~2 of \cite{heg2003}.
This value is also smaller than $R_{h}/R_{j}=0.35$, given by A23. This points out that it is a tidally filled OC.  

When the all evidences of no any signs of expansion/shrinkage (Fig.~15b-c), a high $R_{core}/R_{h}$, compact $R_{h}/R_{t}$ and $R_{h}/R_{j}$ ratios  are evaluated together as the  dynamic indicators, NGC~1513 exposes less to the tidal stripping/disruption of the external perturbations without being tidally disrupted. These external perturbations are the collisions with massive GMCs, tidal effects from spiral arms, disc and Bulge crossings.
Its distant location $R_{GC}=9.57$ kpc has also an advantage in preserving stellar content against these external perturbations.

As emphasized in the paper of \cite{kar2023}, a tidal radius much larger than the Jacobi radius ($R_{t}/R_{J}\gtrsim 1$) might be linked to significant mass loss due to tidal effects. $R_{t}/R_{J}$ is the Roche volume filling factor, as the tidal to Jacobi radius ratio. A23 state that the OCs with $R_{t}/R_{J}\gtrsim 1.25$ are tidally overfilled. The OCs with $R_{t}/R_{J}<1$ keep their stellar content within its Jacobi radius. In light of this information, its Roche lobe is gravitationally bound to the cluster, and thus  NGC~1513 with $R_{t}/R_{J}=0.54$ appears to keep its stellar content with the advantage of exposing less to the tidal effects due to its distant position plus its one revolution around the center of the galaxy. 
$R_{J}=20.20$ pc value given by \cite{hunt2024} supports this conclusion, which is in compatible with our $R_{J}=19.88$ pc value (Col.~2 of Table 9). 
In summary, for NGC~1513,  mass segregation and two body relaxation as internal dynamical evolution seem to be efficient than cluster evaporation process.

We find that NGC~1513 migrated about 0.46~kpc from its birth place. It has guiding radius fairly near its birth radius. Its least migration rate is also compatible with the cluster's low evaporation due to low-mass loss and little  dynamical evolution (small $\tau$).

\section*{Acknowledgments}
We thank Dr. M.~Netopil for providing us the metal abundance value from his DG method, and for the valuable discussion on the analysis. R.~Sagar is also thanked for providing CCD $UBV$ data of NGC~1513 for the comparison.  We also thank Drs. O.~Gunes and M.~Angelo for valuable comments on the dynamical evolution. The open cluster data is based upon observations carried out at the Observatorio Astronómico Nacional on the Sierra San Pedro Mártir (OAN-SPM), Baja California, México.  This paper has made use of results from the European Space Agency (ESA) space mission Gaia, the data from which were processed by the Gaia Data Processing and Analysis Consortium (DPAC). Funding for the DPAC has been provided by national institutions, in particular the institutions participating in the Gaia Multilateral Agreement. The Gaia mission website is http: //www.cosmos.esa.int/gaia. This paper has also made use of the WEBDA database, operated at Department of Theoretical Physics and Astrophysics of the Masaryk University, Brno (the Vienna site is just a backup). This publication also makes use of SIMBAD database-VizieR (http://vizier.u-strasbg.fr/viz-bin/VizieR?-source=II/246.).

%%%%%%%%%%%%%%%%%%%%%%%%%%%%%%%%%%%%%%%%%%%%%%%%%%
\section*{Data Availability}
The photometric data of NGC~1513 can be requested from Raul Michel (rmm@astro.unam.mx).




\end{document}